\documentclass[11pt,letterpaper]{article}
\usepackage{amsmath,amsthm,amssymb,amsfonts}
\usepackage{mathtools}
\mathtoolsset{centercolon}
\usepackage{tikz}
\usepackage[margin=1.20in]{geometry}
\usetikzlibrary{calc,positioning,arrows,shapes.symbols,automata,chains,fit,shapes,er,petri,patterns}

\usepackage{hyperref}

\hypersetup{
  pdftitle = {Quantum hedging in two-round prover-verifier interactions},
  pdfauthor = {Srinivasan Arunachalam, Abel Molina, Vincent Russo}
}

\definecolor{darkred}  {rgb}{0.5,0,0}
\definecolor{darkblue} {rgb}{0,0,0.5}
\definecolor{darkgreen}{rgb}{0,0.5,0}
\hypersetup{
  urlcolor   = blue,         % color of external links
  linkcolor  = darkblue,     % color of internal links
  citecolor  = darkgreen,    % color of links to bibliography
  filecolor  = darkred       % color of file links
}

\newtheorem{theorem}{Theorem}
\newtheorem{lemma}[theorem]{Lemma}

\newtheorem{corollary}[theorem]{Corollary}

\DeclarePairedDelimiter{\abs}{\lvert}{\rvert}
\DeclarePairedDelimiter{\norm}{\lVert}{\rVert}

\newcommand{\floor}[1]{\left\lfloor #1 \right\rfloor}

\renewcommand{\(}{\left(}
\renewcommand{\)}{\right)}
\newcommand{\lket}{\left|}
\newcommand{\rket} {\right\rangle}
\newcommand{\lbra}{\left\langle}
\newcommand{\rbra}{\right|}
\newcommand{\microspace}{\mspace{0.5mu}}
\newcommand{\ket}[1]{\lket\microspace #1 \microspace\rket}
\newcommand{\bra}[1]{\lbra\microspace #1 \microspace\rbra}
\newcommand{\ketbra}[2]{|#1\rangle\langle#2|}

\newcommand{\class}[1]{\textup{#1}}
\newcommand{\reg}[1]{\mathsf{#1}}

\newcommand{\I}{\mathcal{I}}
\newcommand{\X}{\mathcal{X}}
\newcommand{\Y}{\mathcal{Y}}
\newcommand{\Z}{\mathcal{Z}}

\newcommand{\setft}[1]{\mathrm{#1}}
\newcommand{\vecnotation}[1]{\setft{vec}\left(#1\right)}
\newcommand{\trans}[1]{\setft{T}\left(#1\right)}
\newcommand{\pos}[1]{\setft{Pos}\left(#1\right)}
\newcommand{\lin}[1]{\setft{L}\left(#1\right)}
\newcommand{\density}[1]{\setft{D}\left(#1\right)}
\newcommand{\Density}{\setft{D}}
\newcommand{\Pos}{\setft{Pos}}
\newcommand{\Lin}{\setft{L}}
\newcommand{\Channel}{\setft{C}}
\newcommand{\Herm}{\setft{Herm}}

\newcommand{\Trans}{\setft{T}}

\newcommand{\ip}[2]{\left\langle #1 , #2\right\rangle} % inner product
\DeclareMathOperator{\tr}{Tr}

\newcommand{\footremember}[2]{%
   \footnote{#2}
    \newcounter{#1}
    \setcounter{#1}{\value{footnote}}%
}
\newcommand{\footrecall}[1]{%
    \footnotemark[\value{#1}]%
} 

\begin{document}

\title{Quantum hedging in two-round prover-verifier interactions}

%\author{Srinivasan Arunachalam}
%\affiliation{}
%\author{Abel Molina}
%\affiliation{Institute for Quantum Computing and David R. Cheriton School of Computer Science, University of Waterloo, Ontario, Canada}
\author{Srinivasan Arunachalam\footremember{1}{QuSoft, CWI}, Abel Molina\footremember{2}{%
    Institute for Quantum Computing \& School of Computer Science, 
    University of Waterloo.} , and Vincent Russo\footrecall{2}
}

%
% \affiliation{$^3$David R. Cheriton School of Computer Science, University of Waterloo, Ontario, Canada}

\maketitle

%
%%%%%%%%%%%%%%%%%%%%%%
\begin{abstract}
We consider the problem of a particular kind of quantum correlation that arises in some two-party games. In these games, one player is presented with a question they must answer, yielding an outcome of either ``win'' or ``lose''.  Molina and Watrous~\cite{molina2012hedging} studied such a game that exhibited a perfect form of \emph{hedging}, where the risk of losing a first game can completely offset the corresponding risk for a second game. This is a non-classical quantum phenomenon, and establishes the impossibility of performing strong error-reduction for quantum interactive proof systems by parallel repetition, unlike for classical interactive proof systems. We take a step in this article towards a better understanding of the hedging phenomenon by giving a complete characterization of when perfect hedging is  possible for a natural generalization of the game in \cite{molina2012hedging}. Exploring in a different direction the subject of quantum hedging, and motivated by implementation concerns regarding loss-tolerance, we also consider a variation of the protocol where the player who receives the question can choose to restart the game rather than return an answer. We show that in this setting there is no possible hedging for any game played with state spaces corresponding to finite-dimensional complex Euclidean spaces.
\end{abstract}

\section{Overview and motivation}

\label{sec:overviewandmotiv}

The interactions we study consist of parallel repetitions of a game played between players Alice and Bob, also referred to as the {\it verifier} and {\it prover} respectively. The setting of the game is:

\begin{enumerate}
	\item Alice prepares a question, and sends this question to Bob. 
	\item Bob generates an answer,  and sends it back to Alice. 
	\item Alice evaluates this answer and decides if Bob wins or loses.
\end{enumerate}

\noindent It is assumed that Bob has complete knowledge of Alice's specification, including both the method used to determine Alice's question and the criteria that she uses to determine whether Bob has won or lost the game.

Molina and Watrous~\cite{molina2012hedging} consider a specific instance of this setting where Alice sends half of a 2-qubit Bell state $\frac{1}{\sqrt{2}} \ket{00}~+~\frac{1}{\sqrt{2}} \ket{11}$ to Bob. Bob replies with a qubit and Alice evaluates Bob's answer by measuring his qubit and the second half of the Bell state against the state $\cos(\pi/8)\ket{00} +\sin(\pi/8) \ket{11}$. A victory for Bob corresponds to the outcome of Alice measurement corresponding to  $\cos(\pi/8)\ket{00} +\sin(\pi/8) \ket{11}$. When Alice and Bob play two repetitions of this game in parallel, the results in \cite{molina2012hedging} show that there exists a strategy for Bob that guarantees he wins \emph{at least one} of the two repetitions with probability~1. However, when the game is played once, the probability that Bob wins is at most $\cos(\pi/8)^2 \approx 0.8536$. Playing two repetitions in parallel leads then to a \emph{hedging} phenomenon, where if Bob wants to decrease his chance of losing both repetitions, he can do so by not playing each game independently and optimally. This hedging is also \emph{perfect}, in the sense that Bob can completely offset the risk of losing both games.

This is a completely quantum phenomenon, with no classical counterpart. Indeed, when classical information is considered, and for any game that fits the setting we study, it is immediate to show that when Bob wants to win at least $k$ out of $n$ parallel repetitions, it is optimal for him to play independently (however, this is not the case when considering multiple provers \cite{fortnow1989complexity, feige1991success,  raz1998parallel, holenstein2007parallel, braverman2015small}). This establishes the non-triviality of the set of outcome distributions that are possible to obtain from parallel repetition of the games that we study, when compared to the classical case. In particular, it immediately illustrates that the technique of parallel repetition cannot be used to trivially achieve strong error reduction for the complexity class $\class{QIP(2)}$, a class studied for example in \cite{raz2005quantum, wehner2006entanglement, jain2009two, Hayden2014}. The quantum hedging phenomenon is also an example where the quantum version of a game produces outcomes unachievable by its classical counterpart. Most famously considered by Bell \cite{bell1964einstein}, this type of violation has been observed in a number of game-like frameworks~\cite{clauser1969proposed, mermin1990simple, peres1990incompatible,  cleve2004consequences, buhrman2011complexity, regev2012bell,Cooney2015}.

It is natural then to ask how general is the hedging phenomenon, both qualitatively and quantitatively. A complete understanding of this question would allow us to characterize the outcome distributions that can arise from Alice and Bob playing $n$ parallel repetitions of a prover-verifier game in our setting. Consequently, it could lead to a protocol for achieving error reduction via parallel repetition for \class{QIP(2)} simpler than the one currently known~\cite{jain2009two}. The techniques used to achieve such an understanding could conceivably also extend to the analysis of prover-verifier games involving further rounds of communication, and more generally to other kinds of multi-party quantum interactions. This would lead to results for the corresponding complexity classes (and likely also for their classical parallels) about error reduction by parallel repetition. Taking a step towards such a complete understanding, we consider in Section~\ref{sec:general-model} a 2-parameter generalization of the game in~\cite{molina2012hedging}, and characterize when Bob can guarantee that he wins at least $1$ out of $n$ parallel repetitions, for every $n$. We also give optimal strategies for Bob to win at least $1$ out of $n$ parallel repetitions, both when perfect hedging is possible and not possible. We believe these findings are a valuable stepping stone towards a more complete understanding of hedging behaviors for fully arbitrary initial states, fully arbitrary quantum measurements, and $k$-out-of-$n$ settings, as well as highly non-trivial from a mathematical point of view. The formulas that we obtain also open the door for connections between the hedging phenomenon and recent work \cite{Bandyopadhyay2014} involving generalizations of the PBR game  \cite{pusey2012reality}, as we will discuss further in Section \ref{sec:discussion}.

Exploring in a different direction the subject of quantum hedging, it also seems natural  to consider the possibility of implementing a game that exhibits quantum hedging using existing quantum information processing devices. One possible choice would be to use optical quantum devices, but the immediate concern arises  \cite{devin-smith} of how to account for the fact that photon losses will often occur, leading to a communication error between Alice and Bob. Even if one chose another implementation method where communication is more reliable, one would still need to consider the general fact that communication errors can occur. More generally, the consideration of implementation inaccuracies is a standard direction in which to extend results concerning quantum information protocols -- see for example recent work regarding loss-tolerant protocols for quantum coin-flipping \cite{aharon2010family} and QKD, \cite{tamaki2014loss} and noise-tolerant protocols for quantum money \cite{pastawski2012unforgeable}, quantum coin-flipping \cite{zhang2015quantum} and quantum randomness amplification \cite{brandao2016realistic}.

Along this direction, we consider a loss-tolerant formalism in Section~\ref{sec:protocol-errors}, and prove that under our formalism quantum hedging is not possible. To model communication errors, we assume that Alice cannot distinguish a communication error from Bob choosing not to return an answer. Therefore, our formalism simply allows for the possibility that Bob chooses not to return an answer, in which case the game is repeated. Bob choosing in our formalism a random whether to return an answer or not would correspond to a genuine disruption of communication, while Bob strategizing about when to return an answer would correspond to Bob using communication errors as an excuse to avoid a losing outcome. Our particular choice of framework can also be seen as adding postselection to two-round quantum  prover-verifier interactions. This addition of post-selection has been previously considered in the case of single-party quantum computation~\cite{aaronson2005quantum, scegulnaja2010postselection, yakaryilmaz2011probabilistic, mahadev2015rational}, but not to our knowledge in the context of quantum prover-verifier interactions. 

The techniques used to obtain our results in Section~\ref{sec:protocol-errors} are inspired by the techniques in \cite{fiuravsek2004optimal}, which studies a particular case of quantum cloning. The connection between quantum cloning and semidefinite programming was observed in \cite{audenaert2002optimizing}, and has been used to obtain results regarding quantum cloning (see the review in \cite{Cerf2006}). However, this is the first time to our knowledge that this connection with semidefinite programming acts as a bridge to apply ideas about optimal quantum cloning to the context of fully general two-round quantum prover-verifier interactions.

Both of our results leave room for further progress. In particular, one can consider hedging in a wider context than the setting in Section~\ref{sec:general-model}, and consider formalisms that model communication errors in a different way than in Section~\ref{sec:protocol-errors}. We give  some suggestions in Section~\ref{sec:discussion} concerning corresponding choices for further exploration.   

%%%%%%%%%%%%%%%%%%%%%%
\section{Notation} \label{sec:notation}

We will denote the set of binary strings with length $n$ as $\{0, 1\}^n$. These strings will be indexed from $0$ to $n-1$. Therefore, we will denote the $n$ successive binary symbols or bits in $a \in \{0, 1\}^n$ as $a_0, \ldots a_{n-1}$. $\land r, \lor r$, and $\bigoplus r$ refer to the logical AND, OR, and XOR of the bits of $r \in \{0,1\}^n$, respectively, while $\abs{r}$ refers to its Hamming weight.

Vector spaces associated with a quantum system are defined as complex Euclidean spaces. We denote these spaces by the capital script letters $\X, \Y,$ and $\Z$. The dual of a vector $x$ in a complex Euclidean vector space $\X$  will be the linear functional $A : \X \rightarrow \mathbb{C}$ that maps $y$ to $\ip{x}{y}$. For a $d$-dimensional complex Euclidean space, we will often fix a standard \emph{computational} basis and, using bra-ket notation, address its elements and their duals as $\{ \ket{0}, \ldots, \ket{d -1}\}$  and   $\{ \bra{0}, \ldots, \bra{d -1}\}$, respectively. The encoding of the label inside a bra or a ket will often be done in binary for ease of explanation.

The complex vector space of linear operators of the form $A : \X \rightarrow \Y$ is denoted by $\Lin(\X,\Y)$. We write $A \in\Lin(\X)$ as a shorthand for $A : \X \rightarrow \X$. The adjoint $X^*$ of an operator $X \in \Lin(X)$ is the operator such that for all $u, v \in \X$, $\ip{u}{Xv} = \ip{X^*u}{v}$. An operator $H \in \Lin(\X)$ is {\it Hermitian} if $H = H^*$. We write $\Herm(\X)$ to denote the set of all Hermitian operators. The inner product $\ip{A}{B} = \tr(AB)$ between two operators $A,B \in \Herm(\X)$ is real and satisfies $\ip{A}{B} = \ip{B}{A}$. If an operator $P \in \Herm(\X)$, and all eigenvalues of $P$ are non-negative, then we call $P$ {\it positive semidefinite}, and refer to all such operators as $P \in \Pos(\X)$. For a Hermitian operator $H$, $\norm{H}$ denotes the operator norm of $H$, that is, the largest absolute value of an eigenvalue. If for an operator $\rho \in \Pos(\X)$ it is the case that $\tr(\rho) = 1$, then $\rho$ is said to be a \emph{density operator}, and is referred to as $\rho \in \Density(\X)$. We adopt the convention of writing $\I_{\X}$ as opposed to $\I$ to indicate that the identity is acting on the space $\X$ when convenient to do so. We will define the $\setft{vec}: \Lin(\X,\Y) \rightarrow \X \otimes \Y$ mapping to be the one that takes $yx^*$ to $x \otimes y$, for $x$ and $y$ elements of the standard/computational basis of $\X$ and $\Y$. This can be seen as flattening a matrix into a vector. For any two operators $A,B \in \Lin(\X,\Y)$, it will hold that $\ip{A}{B} = \ip{\vecnotation{A}}{\vecnotation{B}}$.

We also consider linear mappings of the form $\Phi~:~\Lin(\X) \rightarrow \Lin(\Y)$. The space of all such mappings is denoted as $\Trans(\X,\Y)$. For each $\Phi \in \trans{\X,\Y}$, a unique adjoint mapping $\Phi^* \in \trans{\Y,\X}$ is defined by the property that $\ip{Y}{\Phi(X)} = \ip{\Phi^*(Y)}{X}$ for all $X \in \Lin(\X)$ $Y \in \Lin(\Y)$. Throughout this work, we define quantum states by the set of density operators $\rho \in \Density(\X)$, with $\X$ a complex Euclidean space. Associated with the space $\X$ one may consider a {\it register} denoted $\reg{X}$ in which the state $\rho$ is contained. We consider measurements of a register $\reg{X}$ as being described by a set of positive semidefinite operators $\{P_a : a \in \Sigma\}$ indexed by a finite non-empty set $\Sigma$ of measurement outcomes which satisfies the constraint $\sum_{a \in \Sigma} P_a = \I_{\X}$.  By performing a measurement on $\reg{X}$ in state $\rho$, the outcome $a \in \Sigma$ results with probability $\ip{P_a}{\rho}$.  These measurements are known as POVMs. We can also consider quantum states stored across $n$ registers $(\reg{X_1}, \reg{X_2}, \cdots, \reg{X_n} )$. We can describe the joint state of those registers by a density operator $\sigma \in \Density (\X_1 \otimes \cdots \otimes \X_n)$.

A linear mapping $\Phi : \Lin(\X) \rightarrow \Lin(\Y)$ is said to be \emph{completely positive} if $\Phi\otimes \I_\Z$ is a map that preserves positive semidefiniteness for every complex Euclidean space $\Z$ and $\Phi$ is said to be \emph{trace-preserving} if $\tr(\Phi(X))=\tr(X)$ for all $X\in L(\X)$. We define a {\it quantum channel} as a linear mapping $\Phi : \Lin(\X) \rightarrow \Lin(\Y)$ that is completely positive and trace preserving. A channel transforms some state $\rho$ stored in register $\reg{X}$ into the state $\Phi(\rho)$ of another register $\reg{Y}$. The set of all channels between such two registers is denoted by $\Channel(\X,\Y)$, and is a compact and convex set. Note that the channel corresponding to an unitary operator $U$ is the one that maps a quantum state $\sigma$ to $U \sigma U^*$.

For spaces $\X$ and $\Y$, one may define the Choi representation of an operator $\Phi \in \Trans(\X,\Y)$ as  $J(\Phi) = \sum_{i,j} \Phi \left( \ket{i}\bra{j}\right) \otimes \ket{i}\bra{j}$, where $J : \Trans(\X,\Y) \rightarrow \Lin(\Y \otimes \X)$, and $i$ and $j$ iterate over the computational basis for $\X$. Note that the mapping $J$ is linear, bijective, and multiplicative with respect to the tensor product. The Choi representation has a number of more complex properties, three of which will be useful to us:
\\
\begin{lemma} \leavevmode
\label{ChoiLemma}
\begin{enumerate}
\item The mapping $\Phi$ is completely positive if and only if $J(\Phi) \in \Pos(\Y \otimes \X)$. 
\item The mapping $\Phi$ is trace preserving if and only if $\tr_{\Y}(J(\Phi)) = \I_{\X}$
\item $\Phi(Z) = \tr_{\X} \left[ J(\Phi) \left(\I_{\Y} \otimes Z^{\text{T}}\right) \right]$
\end{enumerate}
\end{lemma}

\noindent We refer the reader to \cite{Watrous2011} for the proof of Lemma~\ref{ChoiLemma} and further details on the notation.

%%%%%%%%%%%%%%%%%%%%%%
\section{Hedging to win $1$ out of $n$ parallel repetitions of a game} \label{sec:general-model}

Let $G$ denote the following game:

\begin{enumerate}
\item Alice prepares the 2-qubit state $\rho_{\alpha} = u_\alpha u_{\alpha}^* \in \Density(\X \otimes \Z)$ in registers $(\reg{X},\reg{Z})$ where 
	\begin{align} \label{eq:Alicestate}
		u_{\alpha} = \alpha \ket{00} + \sqrt{1 - \alpha^2} \ket{11} \in \X \otimes \Z,
	\end{align}
	for $\alpha \in (0,1]$. Alice sends register $\reg{X}$ to Bob. 
\item Bob applies a channel $\Phi \in \Channel(\X,\Y)$ to the contents of $\reg{X}$. This results in a state $\sigma \in \Density(\Y \otimes \Z)$, contained in registers $(\reg{Y},\reg{Z})$. Register $\reg{Y}$ is sent back to Alice. 
\item Alice performs a measurement on the state $\sigma$. This measurement is $\{ P_{0,\theta}, P_{1,\theta} \}$  for $\theta \in [0,2\pi) $, with \begin{align}
 P_{1,\theta} = v_{\theta} v_{\theta}^*, ~P_{0,\theta} = \I - P_{1,\theta},   \nonumber \\
 v_{\theta} = \cos(\theta) \ket{00} + \sin(\theta) \ket{11} \in \Y \otimes \Z. \label{eq:AliceMeasurement} 
 \end{align}
 
An outcome of ``0'' or ``1'' denotes a losing or winning outcome for Bob, respectively. 
\end{enumerate}

One can imagine repeating the game $G$ $n$ times in parallel. This is denoted as $G^n$, and illustrated in Figure \ref{fig:hedging-n}. In this setting, Alice prepares $n$ states $\rho_{1,\alpha}, \ldots, \rho_{n,\alpha}$ in registers $\left( (\reg{X_1},\reg{Z_1}), \cdots, (\reg{X}_n,\reg{Z}_n) \right)$ where 
\begin{align}
	\rho_{1,\alpha} \in \Density(\X_1 \otimes \Z_1), \ldots, \rho_{n,\alpha} \in \Density(\X_n \otimes \Z_n).
\end{align}
Alice sends the registers $(\reg{X}_1, \ldots, \reg{X}_n)$ to Bob and he applies his quantum channel,
\begin{align} \label{eq:phi-form}
	\Phi_n \in \Channel(\X_1 \otimes \cdots \otimes \X_n, \Y_1 \otimes \cdots \otimes \Y_n).
\end{align} 
The resulting states are sent back to Alice and she performs a series of $n$ projective measurements with respect to the operators $P_{0,\theta}, P_{1,\theta}$. These give $n$ outcomes of either $0$ or $1$, loss or win. Since Bob's actions are not required to respect the independence of the measurements, they may cause correlations between the $n$ measurement outcomes.

Indeed, in \cite{molina2012hedging}, Molina and Watrous analyzed $G^n$ for $n = 2$ where $\alpha = 1/\sqrt{2}$ and $\theta = \pi/8$, and found that Bob wins one out of the two games with certainty if he applies a specific correlated strategy. If on the other hand, Bob treated each repetition independently, it would \emph{not} be guaranteed that Bob would win at least one of the games. 

We consider $G^n$ for any $n \geq 1$ and ask for what values of $\alpha$ and $\theta$ is it true that Bob can make sure to win with certainty \emph{at least} one out of the $n$ games in $G^n$. Let $p_{n,\alpha,\theta}(\Phi_n) \in [0,1]$ be the probability that Bob loses all $n$ outcomes of $G^n$ using the strategy defined by $\Phi_n$. This is given by: 
\begin{align}
	p_{n,\alpha,\theta}(\Phi_n) = \ip{ P_{0,\theta}^{\otimes n} }{ \left( \Phi_n \otimes \I_{\Z_1 \otimes \cdots \otimes \Z_n} \right) \left( \bigotimes_{i=1}^n \rho_{i,\alpha} \right) }. 
\end{align}

Let $m_{n,\alpha,\theta} \in [0,1]$ be $\min_{\Phi_n} p_{n,\alpha,\theta}(\Phi_n)$. We refer to a quantum channel $\Phi_n$ that minimizes $m_{n,\alpha,\theta}$ as an \emph{optimal strategy}. That is, equal to the \emph{minimum probability} with which Bob loses each game over all choices of quantum channels $\Phi_n$ of the form in~\eqref{eq:phi-form}. If $m_{n,\alpha,\theta}$ evaluates to 0, then there exists a $\Phi_n$ that ensures Bob wins at least one~game. 

The quantity $m_{n,\alpha,\theta}$ is expressible as the optimal value of a semidefinite program. Let $Q_{0,\alpha,\theta} \in \Pos(\Y_i \otimes \X_i)$ be defined as 
\begin{align} \label{eq:Q_a}
	Q_{0,\alpha,\theta} = \left( \I_{\Y_i} \otimes \Psi_{\rho_{\alpha}} \right) \left( P_{0,\theta} \right), 
\end{align}
where the mapping $\Psi_{\rho_{\alpha}} : \Lin(\Z) \rightarrow \Lin(\X)$ is defined by $J(\Psi_{\rho_{\alpha}}) = \overline{\rho_{\alpha}}$ (the entry-wise complex conjugate of $\rho_\alpha$). This makes $Q_{0,\alpha,\theta}$ a function of both $ P_{0,\theta}$ and $\rho_{\alpha}$.

It follows from Lemma~1 of  \cite{molina2012hedging} that $Q_0$ is positive semidefinite, and that for any channel $\Phi~:~\Lin(\X)~\rightarrow~\Lin(\Y)$,  we have $\ip{P_{0,\theta}}{ \left( \Phi \otimes \I \right) (\rho_{i, \alpha})}  = \ip{Q_{0,\theta}}{J(\Phi)}$. This can be proved by considering the case where $\rho_{i, \alpha}$ corresponds to a rank-1 operator that transforms a state of the computational basis into another one, and then using the linearity properties of the inner product (see Appendix \ref{app:proof-lemma-mw12}) for more details of this derivation). Putting this together with facts 1 and 2 about the Choi representation in Lemma~\ref{ChoiLemma}, and the bijective property of the $J(\cdot)$ map, we obtain that the following primal and dual pair gives a semidefinite program to compute $m_{n,\alpha,\theta}$:

\begin{figure}[!htbp]
\begin{center}
\begin{tikzpicture}[text depth=1pt, bend angle=45]
	\tikzstyle {Alice} = [shape=rectangle, rounded corners, draw=black!50, thick, fill=black!20, minimum height=9mm, minimum width=9mm, inner sep=1ex]
    \tikzstyle {Bob}   = [shape=rectangle, rounded corners, dotted, draw=black!50, thick,fill=gray!15, minimum height=20mm, minimum width=10mm] 

	% rho_1 ... rho_n
	\node[Alice] (Alice-1-1) at (0,0) {$\rho_{1,\alpha}$};
	\node[Alice] (Alice-1-2) at (0,1.5) {$\rho_{2,\alpha}$};	
	\node[] (elips-1) at (0,2.5) {$\vdots$};
	\node[Alice] (Alice-1-n) at (0,3.3) {$\rho_{n,\alpha}$};

	% P_{a_1} ... P_{a_n}
	\node[Alice] (Alice-2-1) at (6,0) {\small $\{P_{a_1}\}$};
	\node[Alice] (Alice-2-2) at (6,1.5) {\small $\{P_{a_2} \}$};	
	\node[] (elips-2) at (6,2.5) {$\vdots$};
	\node[Alice] (Alice-2-n) at (6,3.3) {\small $\{P_{a_n} \}$};	
	
	% Bob (Phi)
    \node[Bob] (Bob) at (3,5.5) {\large $\Phi_n$};

	\node[] (elips-3) at (0.9,4.4) {$\vdots$};    
    \node[] (elips-4) at (5.1,4.4) {$\vdots$};
	\node[] (elips-5) at (2.9,2.6) {$\vdots$};
       
    \draw[->] (Alice-1-1) -- node[above]{\small $\Z_1$} (Alice-2-1);
    \draw[->] (Alice-1-2) -- node[above]{\small $\Z_2$} (Alice-2-2);
    \draw[->] (Alice-1-n) -- node[above]{\small $\Z_n$} (Alice-2-n);
       
   	\draw[->] (Alice-1-1) edge[->, bend left=0] node[above left]{\small $\X_1$} (Bob); 
   	\draw[->] (Alice-1-2) edge[->, bend left=10] node[above left]{\small $\X_2$} (Bob); 
   	\draw[->] (Alice-1-n) edge[->, bend left=30] node[above left]{\small $\X_n$} (Bob); 
   	
   	\draw[->] (Bob) edge[->, bend left=0] node[above right]{\small $\Y_1$} (Alice-2-1); 
   	\draw[->] (Bob) edge[->, bend left=10] node[above right]{\small $\Y_2$} (Alice-2-2); 
   	\draw[->] (Bob) edge[->, bend left=30] node[above right]{\small $\Y_n$} (Alice-2-n); 
\end{tikzpicture}
\end{center}
\caption{The parallel repetition $G^n$ of $n$ copies of a game $G$ of the type we study.}
\label{fig:hedging-n}
\end{figure}
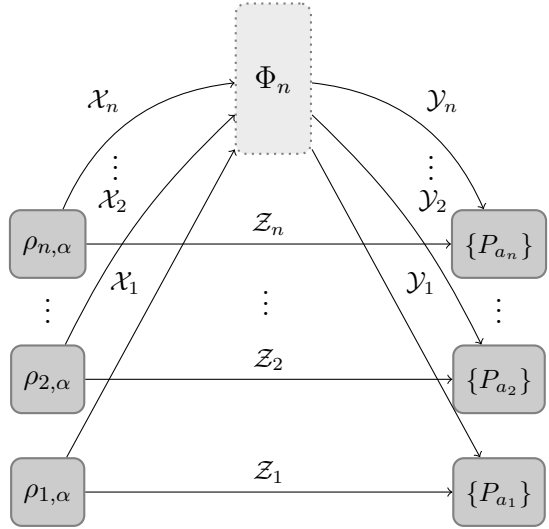

\newpage

\begin{center}
    \centerline{\underline{$m_{n,\alpha,\theta}$: Primal problem}}\vspace{-5mm}
    \begin{equation}
   \label{eq:1OutOfNSDP}
    \begin{aligned}
      \text{minimize:}\quad & \ip{Q_{0,\alpha,\theta}^{\otimes n}}{X} \\
      \text{subject to:}\quad & \tr_{\Y_1 \otimes \cdots \otimes \Y_n}(X) = \I_{\X_1 \otimes \cdots \otimes \X_n},\\
      & X \in \Pos(X_1 \otimes \Y_1 \otimes \cdots \otimes \X_n \otimes \Y_n).
    \end{aligned}
    \end{equation}
\end{center}

\begin{center}
    \centerline{\underline{$m_{n,\alpha,\theta}$: Dual problem}}\vspace{-5mm}
    \begin{equation}
   \label{eq:1OutOfNSDPDual}
    \begin{aligned}
      \text{maximize:}\quad & \tr(Y) \\
      \text{subject to:}\quad & \pi \left(\I_{\Y_1 \otimes \cdots \otimes \Y_n} \otimes Y\right)\pi^{*} \leq Q_{0,\alpha,\theta}^{\otimes n},\\
      & Y \in \Herm(\X_1 \otimes \cdots \otimes \X_n).
    \end{aligned}
    \end{equation}
\end{center}
where $\pi$ is a unitary permutation operator defined by the action
\begin{align*}
	\pi(y_1 \otimes \cdots \otimes y_n \otimes x_1 \otimes \cdots \otimes x_n) = y_1 \otimes x_1 \otimes \cdots \otimes y_n \otimes x_n
\end{align*}
for all $y_1 \in \Y_1, \cdots, y_n \in \Y_n$ and $x_1 \in \X_1, \cdots, x_n \in \X_n$. Note that strong duality holds for the above semidefinite program, by choosing the primal and dual feasible solutions $(X,Y)$ for the application of Slater's theorem as a scalar multiple of the identity. The derivation to obtain this semidefinite program is similar to that in \cite{molina2012hedging}, and previously in \cite{gutoski2007toward} and \cite{gutoski2010quantum}. We point the reader to \cite{code} for MATLAB code that solves SDPs \eqref{eq:1OutOfNSDP} and \eqref{eq:1OutOfNSDPDual}, using the CVX convex optimization package \cite{grant2008cvx}.

We present now for fixed $n$ and $\alpha$ the range of  $\theta$ which characterizes the measurements for which Bob can make sure he wins at least $1$ parallel repetition in $G^n$. That is, it characterizes when is Bob able to perform perfect hedging. Furthermore, we present strategies that give Bob an optimal probability to win at least 1 out of $n$ games, both when Bob is able to perform perfect hedging and when he is not.

\begin{theorem}
	\label{thm:AngleThm}

Let 
\begin{equation}
\label{eq:Anglefor1/n}
		\begin{aligned}
		 \theta_{n,\alpha} &=& \tan^{-1} \left( \sqrt{\frac{1}{\alpha^2} - 1} \left( 2^{1/n} - 1 \right)  \right), \\ 
			 \gamma_{n,\alpha} &=& \tan^{-1} \left( \sqrt{\frac{1}{\alpha^2} - 1} \left( \frac{1}{2^{1/n} - 1} \right) \right),
		\end{aligned}
			\end{equation}
			where the trigonometric domain is restricted to $[0,\pi/2]$. If and only if Alice's rank-1 projective measurement $\{P_0,P_1\}$ is parametrized by $\theta \in \left[\theta_{n,\alpha}, \gamma_{n,\alpha} \right]$, then there exists a strategy for Bob to perform perfect hedging.
			 \end{theorem}

\begin{figure}[!htpb] \label{fig:hedging-plot}
	\begin{center}
		\includegraphics[scale=0.28]{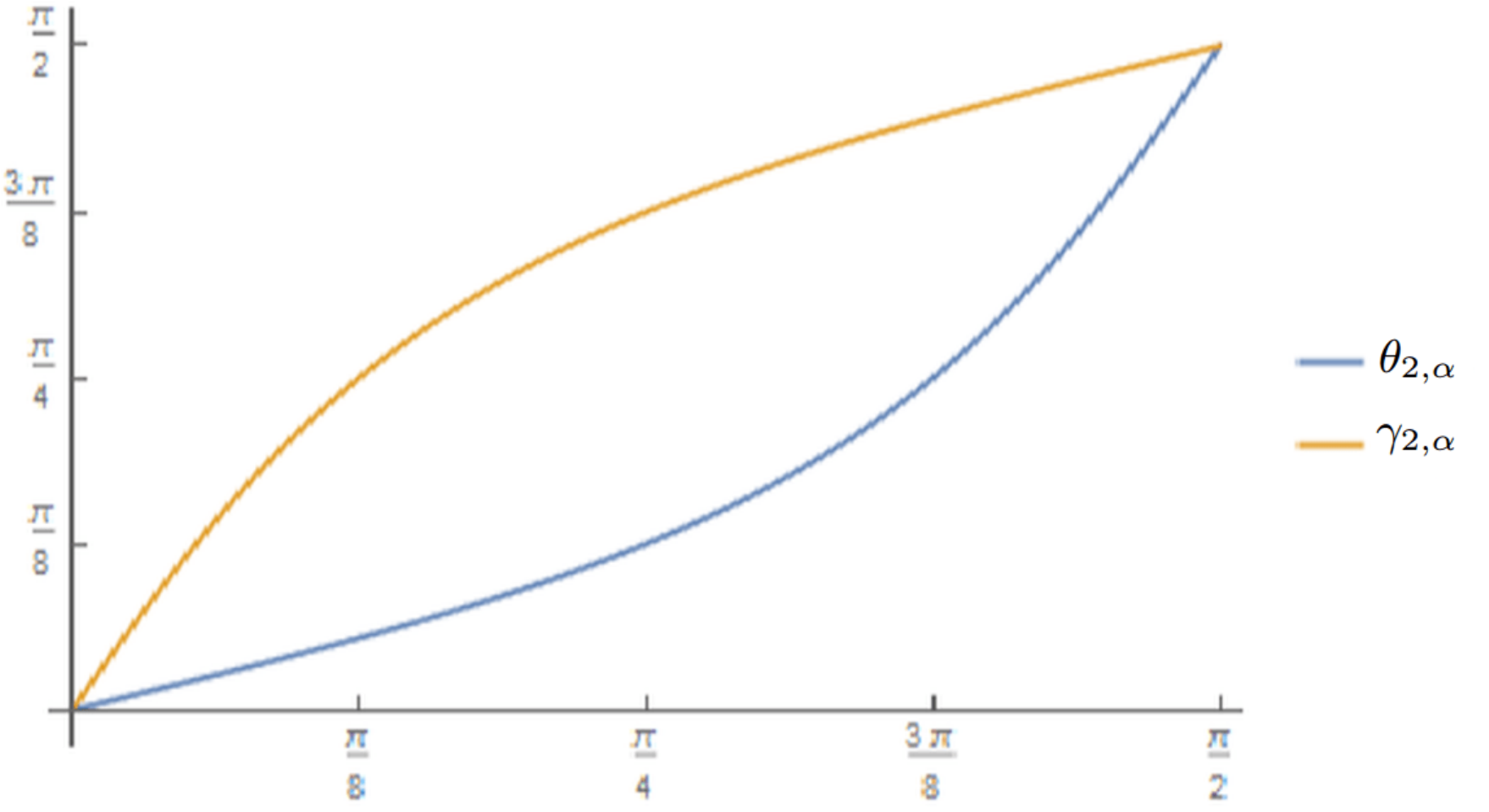}
	\end{center}
		\caption{$\gamma_{2,\alpha}$ and  $\theta_{2, \alpha}$ as a function of $\tan^{-1}\left(\dfrac{\sqrt{1 - \alpha^2}}{\alpha}\right)$}
\end{figure}

We see then that the angle $\pi/8$ used for $\theta$ in \cite{molina2012hedging} corresponds to the lower bound  $\theta_{2, 1/\sqrt{2}} = \pi/8$ from Theorem \ref{thm:AngleThm}, but also that perfect hedging can be attained for this setting up to $\gamma_{2,\frac{1}{\sqrt{2}}} = 3\pi/8$. Note that as the number of games $n$ increases, the size of this range increases. Moreover, for any choice of $\theta$ in $(0, \pi/2)$, there is an $n$ large enough for perfect hedging to be possible. As one can see in our plot of $\theta_{n, \alpha}$ and $\gamma_{n,\alpha}$, the cases where perfect hedging are posssible are symmetric with respect to the case where the initial state and the desired final state are the same (i.e., $\theta = \tan^{-1}(\sqrt{1 - \alpha^2}/\alpha))$. Note also that the size of the range where perfect hedging is possible is minimized for for $\theta=0$ and $\theta=\pi/2$, which correspond to a standard basis measurement done by Alice.

The proof of Theorem \ref{thm:AngleThm} follows immediately from Lemma~\ref{lem:UnitaryStratLem} and Lemma~\ref{thm:optimalstrategy}, stated below. Theorem \ref{thm:AngleThm} results in the following corollary:

\begin{corollary}
\label{cor:maxentangled}
	For a fixed $n$, perfect hedging occurs for the largest range of $\theta$ angles when Alice initially prepares a maximally entangled state (that is, when $\alpha=\frac{1}{\sqrt{2}}$).
\end{corollary}

The proof for the corollary follows from directly maximizing $\gamma_{n,\alpha} - \theta_{n,\alpha}$ over all $\alpha$, by taking derivatives with respect to $\alpha$. This corollary tells us then that an assumption that the initial state is maximally entangled represents an extremal case in our quantum hedging context. One might be able to use when trying to generalize our results, as we will further discuss in Section \ref{sec:discussion}.

In the following lemmas, we define an optimal choice for Bob of the channel $\Phi$ that he applies to the part of the state he receives from Alice:
\begin{lemma}
	\label{lem:StratLem}
	Let $n \geq 2$ be a positive integer, let $\alpha \in (0,1]$, let $\theta_{n,\alpha}$ and $\gamma_{n,\alpha}$ be angles defined as in Theorem~\ref{thm:AngleThm}, and let 
	\begin{equation}
		\label{eq:strategiesOnBorder}
 		\begin{aligned}
			\Lambda_{n} & = & \sum_{r \in \{0,1\}^n} \left(-1\right)^{\land r + \oplus r} \ket{r}\bra{r}, \\
			\Xi_{n} & = & \sum_{r \in \{0,1\}^n} \left(-1\right)^{\lor r + \oplus r} \ket{r}\bra{r}, 
		\end{aligned}
	\end{equation}
	be unitary operators that Bob applies as his strategy in $G^n$. Then it holds that 
	\begin{align}
		p_{n,\alpha,\theta_{n,\alpha}}\left(\Lambda_n\right) = 0 = p_{n,\alpha,\gamma_{n,\alpha}}\left(\Xi_n\right).
	\end{align}
	
\end{lemma}

This shows the existence of strategies $\{\Lambda_n,\Xi_n\}$ for Bob at $\{\theta_{n, \alpha},\gamma_{n, \alpha}\}$ that achieve a value of $0$ for the SDP~\eqref{eq:1OutOfNSDP}. The next lemma proves that for all points within these two bounds there exists such a strategy as well. Note that  $\Lambda_n$  and $\Xi_n$ do not depend on $\alpha$. Also, note that when $n = 2$, Bob's unitary $\Lambda_2$ on the two qubits that he receives is
\begin{align}
	 \Lambda_2 = \begin{pmatrix} 1 & 0 & 0 & 0 \\ 
	                       0 & -1 & 0 & 0 \\ 
	                       0 & 0 & -1 & 0 \\ 
	                       0 & 0 & 0 & -1 
	       \end{pmatrix},
\end{align}
 which gives us the same strategy as in \cite{molina2012hedging}. The proof of the lemma follows from observing that the final state after Bob applies $\Lambda_{n} $/ $\Xi_{n}$ has zero overlap with the state corresponding to Bob losing all the repetitions. The details of the derivation are included in Appendix \ref{app:proof-lemma-strat}.

\begin{lemma}
\label{lem:UnitaryStratLem}
In the scenario where the projective measurements are parametrized by $\theta \in \left[ \theta_{n,\alpha} , \gamma_{n,\alpha}  \right]$ for $\theta_{n,\alpha}$ and $\gamma_{n,\alpha} $ defined as in Theorem \ref{thm:AngleThm}, Bob can apply the strategy corresponding to the following unitary operator to achieve perfect hedging for 1 out of $n$ games:

\textsc{
\begin{equation}
\label{eq:PhiMatrixInside}
\begin{aligned}
(-1)^n |0^n\rangle \langle 0^n| - |1^n\rangle \langle1^n|  +\sum_{i=1}^{n-1} \sum_{\substack{r \in \{0,1\}^n \\ \abs{r}=i}}(-1)^{n+i}k_{r}\ket{r}\bra{r},
\end{aligned}
\end{equation}}

\noindent where for a fixed choice of $|r|=i$, the corresponding $k_{r}$ are ${n \choose i}$ complex numbers with the following properties
\[k_{r}= \begin{cases} 
      s_{\theta, \alpha, n} + i\sqrt{1-s_{\theta, \alpha, n}^2}  \text{ for } \floor{{n \choose i} / 2} \text{values of }r, \\
      s_{\theta, \alpha, n} - i\sqrt{1-s_{\theta, \alpha, n}^2}  \text{ for } \floor{{n \choose i} / 2} \text{values of }r,   \\
      -1 \text{ for the remaining values of }r \text{ when }  {n \choose i} \text{ is }\\
        \text{ odd and }\tan(\theta) \geq \sqrt{\frac{1}{\alpha^2} -1},\\
      1  \text{ for the remaining values of }r \text{ when }  {n \choose i} \text{ is odd } \\
      	 \text{ and }\tan(\theta) < \sqrt{\frac{1}{\alpha^2} -1},\\
   \end{cases}
\]

\noindent where $s_{\theta, \alpha, n}$ is a real number $\in [-1,1]$ whose existence we guarantee in the proof of this lemma.
 
\end{lemma}

Since Bob has complete knowledge of the game, for any $\theta \in \left[ \theta_{n,\alpha} , \gamma_{n,\alpha}  \right]$ Bob can apply the strategy corresponding to the angle $\theta$ selected by Alice. It is clear that the optimal  strategy for Bob is not unique, since our definition does not uniquely specify which coefficients $k_{r}$ correspond to which values of $r$. This lemma is derived by performing a computation (similar to the one for Lemma~\ref{lem:StratLem}) that computes the overlap between the resulting state after Bob applies the strategy we describe and the state corresponding to Bob losing all $n$ repetitions. Then, we consider the cases $s_{\theta, \alpha, n}=-1$  and $s_{\theta, \alpha, n}=1$ and obtain through continuity arguments that there must be a value of  $s_{\theta, \alpha, n}$ in the $[-1,1]$ range that results in perfect hedging.  The details of the corresponding derivation are included in Appendix \ref{app:proof-lemma-unitary}.

We have thus far considered the case when perfect hedging is possible. The following result deals with characterizing the scenario when perfect hedging is not possible, and provides a corresponding strategy for Bob to play~optimally.

\begin{lemma}
\label{thm:optimalstrategy}
For $n \geq 2$ and for $ \theta \in [0, \theta_{n, \alpha}) \cup (\gamma_{n, \alpha}, \pi/2] $ perfect hedging cannot occur, and the strategies $\Lambda_n$ and $\Xi_n$  mentioned in Lemma~\ref{lem:StratLem} are respective optimal strategies for Bob.
\end{lemma}

The proof of this lemma is obtained by using SDP complementary slackness \cite{cs766solutions} to obtain a candidate solution for the dual SDP \eqref{eq:1OutOfNSDPDual} with the same objective value as the  chance of achieving $1$-out-of-$n$ hedging for $\Lambda_n$/$\Xi_n$. Then, one can use a direct sum decomposition of the matrices involved in the SDP constraint to prove the feasibility of this candidate solution. The details of the corresponding calculations are available in Appendix \ref{app:proof-theorem-outside}. Note that the strategy Bob adopts is independent of the parameter $\theta$, implying that when perfect hedging is not possible the strategy is optimal regardless of the projective measurements chosen by Alice. 

It can also be observed from Lemma~\ref{lem:UnitaryStratLem} and Lemma~\ref{thm:optimalstrategy} that a unitary (and in fact, a diagonal in the computational basis) strategy is always sufficient for Bob to win at least once with optimal probability. Note that it intuitively makes sense that Bob's strategy is a diagonal unitary, since switching a $\ket{0}$ to a $\ket{1}$ or vice-versa on his side will produce a state with no overlap with the target state $\cos(\theta)\ket{00} + \sin(\theta)\ket{11}$.

\section{(Lack of) Hedging in a Loss-Tolerant Prover-Verifier Model} \label{sec:protocol-errors}

We consider a variation of the prover-verifier setting where Bob has the choice to not respond to Alice, in  order to model communication errors, as described in Section \ref{sec:overviewandmotiv}. If Bob chooses not to respond, and therefore Alice does not receive an answer, the game is repeated again, and this goes on until an answer is returned by Bob.  Bob might want to do this whenever using his complete knowledge of the game, he can predict  that an answer will result in Alice obtaining a negative outcome in her measurement. Indeed, to see how this variation can change the result of an interaction, consider the following game where Bob is always forced to return an answer:

\begin{enumerate}
\item Alice prepares the maximally entangled state $\frac{1}{\sqrt{2}} \ket{00} + \frac{1}{\sqrt{2}} \ket{11}$ and sends the second qubit to Bob.

\item Bob responds by sending a qubit to Alice.

\item Alice ignores Bob's answer, and measures the qubit she kept with respect to the projective
measurement $\{P_0,P_1\}$, where $P_0 =\ket{1} \bra{1}$ and
$P_1 = \ket{0} \bra{0}$.
\end{enumerate}

It is clear that the maximum probability for Bob to win the game is 50\%. This follows from the fact that the actions of Bob cannot alter the reduced state that Alice holds, and the outcome of the interaction depends only on this state. However, the situation changes drastically when Bob is allowed to return no answer in the second step. In that case, Bob can choose to perform a measurement using the computational basis on the qubit he receives. If the measurement results in the outcome $\ket{0}$, corresponding to $P_1$, he will return an answer, and otherwise he will not, and force a restart. The entanglement between the qubit that Alice keeps and the one that Bob receives guarantees then that the outcome will always be the successful one.

It seems clear then that giving Bob the choice to abort the protocol can have significant changes on what optimal behaviors for Bob are like. This motivates the consideration of whether any form of quantum hedging (perfect or not) is still possible in the  ``repetition after communication error"   setting for an arbitrary two-message quantum-verifier interaction (described by an arbitrary finite-dimensional inital quantum state $\rho$ prepared by Alice and an arbitrary finite-dimensional POVM $\{P_i\}$ used to determine the interaction's outcome.) We ask in this context then whether it will be optimal for Bob to play each interaction independently when trying to optimize his chance of winning at least $k$ out of $n$ parallel interactions.

To answer this question, we will assume in our analysis that Bob always has a nonzero chance of winning a single interaction. If this were not the case, the question of whether or not hedging occurs would be uninteresting. This is because in this case, the optimal probability for Bob to win $k$ out of $n$ parallel repetitions would always be zero. To see why, assume to the contrary that Bob can manage to win $k > 0$ out of $n > 1$ repetitions with non-zero probability. Then, whenever Bob plays a single game with Alice, he could simulate the input for $n-1$ additional interactions, and since the possibility that he wins $k > 0$ of the $n$ games is greater than zero, and the situation is symmetrical, the possibility that he wins the single ``real'' game is greater than zero as well, which contradicts our starting premise.

Furthermore, we need to specify how does the ``repetition after communication error"  aspect of the framework interacts with the ``repeating $n$ interactions in parallel"  aspect of the framework. For simplicity, we will make in our model the assumption that whenever Alice does not receive an answer to one out of $n$ parallel interactions, she will restart all of the $n$ parallel interactions.

To start our analysis, we consider an intermediate setting where we allow Bob to not give an answer, and Alice does not repeat the interaction when she doesn't obtain an answer, and instead counts that as a loss for Bob. This means that Bob can return a state with trace less than one. Using the properties of the Choi representation, and following the same analysis as in \cite{molina2012hedging} and Section~\ref{sec:general-model}, the optimal probability for Bob of achieving outcome $a$ is the value of

\begin{center}
    \centerline{\underline{Primal problem}}\vspace{-5mm}
    \begin{equation}
    \label{sdp-primal}
    \begin{aligned}
      \text{maximize:}\quad & \ip{Q_a}{X}\\
      \text{subject to:}\quad & \tr_{\Y}(X) \leq \I_{\X},\\
      & X\in\pos{\Y\otimes\X},
    \end{aligned}
    \end{equation}
\end{center}

\noindent where $Q_a$ is defined as in ~\eqref{eq:Q_a}, starting from an arbitrary POVM \{$P_i$\} and a state $\rho$. Without loss of generality, we assume that Bob wants to achieve quantum hedging with respect to outcome $a$, and group all other outcomes into a single outcome corresponding to $Q_{1-a}$.

Now we take into account the fact that the interaction is repeated whenever an answer is not received. To do this, it is enough to divide the objective function, which corresponds to the probability of obtaining outcome $a$, by the probability that an answer is returned. This is because we can ignore previous rounds of the interaction, since the repeated rounds occur in series, and Alice acts independently between them. Indeed, the way in which previous rounds would be taken into account would be with an additional input for Bob, corresponding to his memory after the previous rounds of the protocol. But the fact that there is no computational restriction on Bob and no hidden information means that for any possible value of that input, Bob could just simulate the previous rounds to generate it, so the additional memory input is not needed, and we can ignore previous rounds.

Note that the division by the probability that Bob returns an answer would not be possible if Bob just chose not to return an answer. However, that strategy can just be ignored as a non-optimal one, since we are assuming Bob can win with non-zero probability.

The probability that an answer is returned is the trace of the state after Bob returns an answer, which is a linear function of the variable $X$ in SDP \eqref{sdp-primal}. In particular,  the probability is given by $\ip{E}{X}$, where 
\begin{equation}
    \begin{aligned}
      E & = \sum_i Q_i =  \sum_i \left( \I_{\lin{\Y}} \otimes \Psi_{\rho} \right) (P_i) \\
       & =  \left( \I_{\lin{\Y}} \otimes \Psi_{\rho} \right) \I_{\Y \otimes \Z}  =  \I_{\Y} \otimes \tr_{\Z}(\overline{\rho}),
    \end{aligned}
\end{equation}

\noindent and the last step uses the third fact in Lemma~\ref{ChoiLemma}. Note that since $\sum_i Q_i  = E$,  $Q_a \leq E$.

This tells us then how to modify the SDP \eqref{sdp-primal}  that describes Bob's optimal probability of obtaining  outcome $a$ in a way that takes into account our loss-tolerant framework.  In particular, we have that the equivalent of SDP  \eqref{sdp-primal} is now given by 

\begin{center}
    \centerline{\underline{Primal problem}}\vspace{-5mm}
    \begin{equation}
	\label{sdp:finalfinal}
    \begin{aligned}
      \text{maximize:}\quad & \frac{\ip{Q_a}{X}}{\ip{E}{X}}\\
      \text{subject to:}\quad & \tr_{\Y}(X) \leq \I_{\X},\\
      & X\in\pos{\Y\otimes\X}, \ip{E}{X} \neq 0.
    \end{aligned}
    \end{equation}
\end{center}

We use now an analysis inspired by the one in \cite{Cerf2006} to obtain a more explicit form for the value of this SDP. First, notice that scaling a solution $X$ by a nonzero constant will not change the value of the objective function. Since the partial trace operation preserves positive semidefiniteness, we can then get rid of the $\tr_{\Y}(X) \leq \I_{\X}$ constraint:

\begin{center}
    \centerline{\underline{Primal problem}}\vspace{-5mm}
    \begin{equation}
    \begin{aligned}
      \text{maximize:}\quad & \frac{\ip{Q_a}{X}}{\ip{E}{X}}\\
      \text{subject to:} \quad & X\in\pos{\Y\otimes\X}, \ip{E}{X} \neq 0.
    \end{aligned}
    \end{equation}
\end{center}

At this point, we can assume that $X$ corresponds to a rank-one operator. To see why, consider an $X$ that corresponds to a sum of two solutions, $X_1$ and $X_2$. Then, the value of the objective function will be  
\begin{align}
 \frac{\ip{Q_a}{X_1} + \ip{Q_a}{X_2}  }{\ip{E}{X_1} + \ip{E}{X_2}} \leq \max \left(  \frac{\ip{Q_a}{X_1}}{\ip{E}{X_1}},  \frac{\ip{Q_a}{X_2}}{\ip{E}{X_2}} \right),
 \end{align}

\noindent where the inequality follows from the fact that all values on the left-hand side are positive. We obtain the problem

\begin{center}
    \centerline{\underline{Primal problem}}\vspace{-7mm}
    \begin{equation}
    \begin{aligned}
      \text{maximize:}\quad & \frac{x^{\ast}Q_ax}{x^{\ast}Ex}\\
      \text{subject to:} \quad & x \in \Y\otimes\X, x^{\ast}Ex \neq 0.
    \end{aligned}
    \end{equation}
\end{center}

Note now that we can assume without loss of generality that an optimal solution $x$ is contained within the support of $E$.  In this domain the Moore-Penrose pseudo-inverse  of $E$, $E^{+}$, acts as a bijection. Therefore, we replace $x$ by $(E^{+})^{1/2}x$ in the objective function, and obtain

\begin{center}
    \centerline{\underline{Primal problem}}\vspace{-5mm}
    \begin{equation}
    \begin{aligned}
      \text{maximize:}\quad & \frac{x^{\ast}(E^{+})^{1/2}Q_a(E^{+})^{1/2}x}{x^{\ast}x}\\
      \text{subject to:} \quad & x \in \Y\otimes\X, x \perp \ker(E),
    \end{aligned}
    \end{equation}
\end{center}

\noindent which has the value $\norm{(E^+)^{1/2}Q_a(E^+)^{1/2}}$. We denote this as $\norm{\Lambda}$.

When Bob wants to be successful in at least $k$ out of $n$ parallel interactions, with Alice acting independently, one just needs to replace $Q_a$ by the sum of tensor products of $Q_i$'s corresponding to at least $k$ outcomes equal to $a$. Remembering that the sum of all the $Q_i$ is equal to $E$, the same analysis that we performed for a single repetition gives us an optimal probability of $\norm{\Lambda_{k,n}}$, with  $\Lambda_{k,n}$ given by :

\begin{equation}
\begin{aligned}
\Big \lVert (\sqrt{E^+})^{\otimes n} \( E^{\otimes n} - \sum^{k-1}_{t=0} \pi_t \( Q_{1-a}^{\otimes n - t} \otimes Q_{a}^{\otimes t} \) \)  (\sqrt{E^+})^{\otimes n}  \Big \rVert
\end{aligned}
\end{equation}

\noindent where $\pi_t(x)$ is the sum of all $\binom{n}{t}$ unique permutations of $x$.

As an aside, note that one can assume that $\rho$ corresponds to a pure state $\psi$.  This is because given a protocol where Alice initially prepares a mixed state, we can easily modify it so that Alice prepares a purification of that state instead, and just ignores the extra qubits when performing the final measurement. Using this, we observe an interesting fact about this model, which is that at least when one restricts Bob to perform a rank-one measurement, the optimal success probability for Bob does not depend on the Schmidt coefficients of $\psi$. This is proved by letting the initial state that Alice holds be given by $\sum_i \sqrt{p_i} a_i \otimes b_i$, and the state corresponding to Bob's projection by  $\sum_i \sqrt{q_i} c_i \otimes d_i$. Using  algebraic manipulations we obtain that the optimal probability of winning for Bob in a single parallel repetition is
		
\begin{align}
 \Big \lVert \sum_{i,j,k,l} \sqrt{q_j q_l}  b_i^* d_l d_j^* b_k  \overline{a_i} \overline{a_k}^* \otimes c_j c_l^* \Big \rVert,
 \end{align}

\noindent with no dependence on the $p_i$.

This suggests that the example we gave at the beginning of this section might capture all the additional power Bob has in this model. In particular, it suggests that an optimal strategy for Bob might always consist of performing an orthogonal measurement on the qubits he is given, and then refusing to give an answer except when he obtains the ``best'' outcome.

As for our main subject of concern (quantum hedging), it turns out that in the model we just described quantum hedging is not possible. One can interpret this as saying that Bob is already so powerful in one single repetition (since he can choose not to return an answer) than the power to entangle several answers does not add anything in comparison. More precisely, we have the following theorem:

\begin{theorem}
\label{thm:hedging-under-error}
Consider a two-message prover-verifier interaction characterized by an arbitrary initial state $\rho$ and an arbitrary POVM  $\{P_i\}$, both on a finite number of qubits. Then, under the loss-tolerant setting described in this section, it is optimal for Bob to play independently in order to maximize his chance of winning at least $k$ out of $n$ parallel interactions.
\end{theorem}

The statement of the theorem results from a straightforward spectral analysis of the $\Lambda_{k,n}$ operator by induction on $n$ and then $k$. The details of the corresponding computation are included in Appendix \ref{app:proof-theorem-under-error}.

\section{Discussion} \label{sec:discussion}

We have analyzed generalizations of a specific prover-verifier interaction where the verifier can use a quantum hedging strategy to win at least one of $n$ parallel repetitions with a higher probability than what would have been possible playing each game independently. This interesting phenomenon was originally described in \cite{molina2012hedging}, where the authors illustrated an explicit example of perfect hedging when two repetitions of the game were carried out.  It was previously unknown how the perfect hedging phenomenon generalizes to the case when $n$ repetitions of the game are performed. We resolved this question for a generalization of the game in \cite{molina2012hedging}, and provided strategies for Bob that allow him to achieve perfect hedging whenever it is possible.

We also analyzed a variant of this setting where Bob is not obligated to return an answer to Alice. In a practical sense, Bob's refusal to respond to Alice can be viewed in terms of an experimental setup where the lack of a response could correspond to a communication error \cite{devin-smith}. This consideration led to a different semidefinite program that characterized the interaction between Alice and Bob. We then used this SDP 	\eqref{sdp:finalfinal} to ask whether or not Bob still had the ability to take advantage of hedging behavior, with a negative answer. 

While we have considered this hedging behavior in a number of settings, there are still many questions remaining. As mentioned, we have characterized the conditions that allow Bob to win 1 out of $n$ repetitions in a framework that generalizes the game in \cite{molina2012hedging}. However, it still remains open to determine the conditions under which Bob can always win at least $k$ out of $n$ repetitions for some $k > 1$. It would be interesting to determine the threshold of $k$ for which perfect hedging occurs, and to also provide a characterization in regards to the strategy that Bob uses to achieve this result. Running numerical instances for higher values of $k$ and $n$ using a simple formulation in CVX~\cite{grant2008cvx} quickly becomes computationally infeasible, as can be observed from the software we have provided in \cite{code}. It is possible that this code could be optimized to consider further cases,  leading to conjectures regarding the behavior for arbitrary $k$ and $n$ that could be then proved analytically. Based on our current numerical evidence, it is possible that Bob cannot perfectly hedge more than $k = n/2$ games. Note also that when $k \leq n/2$ one can design a strategy for the goal of winning $k$ out of $n$ repetitions by dividing the $n$ parallel repetitions into several smaller groups, and then using the strategies described in this paper in order to always win at least one repetition in each group. It is left as an open question (whose solution we believe to be a significant task) whether the range of parameters in which the resulting strategy always wins $k$ out of $n$ repetitions is the optimal one. Motivated by our results in Corollary \ref{cor:maxentangled}, one could also look into the subject of reducibility between different games in our framework, asking for example whether there is a procedure with an intuitive operational description that transforms a game with an arbitrary shared initial state between Alice and Bob to one where the initial shared state is now maximally entangled, while the possibilities of achieving $k$-out-of-$n$ hedging remains the same.

It is also worth noting that the problem of conclusive state exclusion, which was recently considered in \cite{Bandyopadhyay2014}, seems to be connected to the interaction we have analyzed in this work. In this problem, Alice prepares a mixed state from a given distribution and sends it to Bob, and for Bob to win, he has to accurately discard at least one of the possible options. In \cite{Bandyopadhyay2014} the PBR game, originally formulated in \cite{pusey2012reality}, was analyzed in terms of an semidefinite program using the conclusive state exclusion framework. Some of the formulas we obtain in Section \ref{sec:general-model} are similar to the ones \cite{Bandyopadhyay2014} derive in their analysis of the PBR game, specifically equations \eqref{eq:Anglefor1/n}~and~\eqref{eq:strategiesOnBorder}. Looking at the SDPs involved in their work and in ours, it seems clear that the similarity arises from the fact that diagonal unitaries happen to be optimal for hedging. The fact that they are optimal means that the optimization problem we examine in SDP \eqref{eq:1OutOfNSDP} is equivalent to that of optimizing along complex vectors where each entry of the vector is a unit. Then, to establish the connection with the PBR setting, one would  establish an equivalence between these types of vectors and  highly symmetrical projective measurements like those obtained as optimal solutions in the corresponding PBR state exclusion setting. However, in a setting with initial states outside the $\alpha\ket{00} + \sqrt{1 - \alpha^2}\ket{11}$ family we consider in Section \ref{sec:general-model}, there is no reason why the optimal channel for winning $1$ out of $n$ parallel interactions should correspond to a diagonal unitary. It remains then to see whether any similar connections can be established between such a setting and a state exclusion setting.  It seems plausible that further work clarifying these connections could be used to apply existing results concerning the conclusive state exclusion framework to the hedging framework, and vice versa. 

One could also further consider the setting in which protocol errors are considered. Here, we have assumed that Bob can delay returning an answer for as many iterations of the protocol as he desires. An obvious follow-up question then is to determine whether an advantage from hedging behavior is possible when this is not the case. One might restrain Bob to behaviors where on average he will return an answer within a fixed number of iterations, or introduce constraints be of the form ``After X iterations, Bob's probability of having return an answer must be at least equal to Y''. A special case of those constraints that might be particularly interesting is when Bob is required to return an answer within a fixed number of iterations. We could also modify the way in which the ``repeating after failure" and ``repeating in parallel" frameworks interact. In particular, we could have Alice repeat only a subset of interactions if answers corresponding to the other interactions have been obtained from Bob.

Note that when trying to analyze more general models (in both the ideal and loss-tolerant cases) along the lines described in this section, it might be fruitful to look into whether it is possible to again use ideas from the quantum cloning literature, as we did here in Section \ref{sec:protocol-errors}.  It is possible as well that progress can be made using representation theory tools to simplify or avoid the analysis of semidefinite programs, as done for example in \cite{gatermann2004symmetry, childs2007quantum, christ2010, krovi2015optimal}.

\section*{Acknowledgements}
We would like to thank Devin Smith for the question that led to Section~\ref{sec:protocol-errors} in this paper. A significant amount of thanks is also due to John Watrous for numerous insightful discussions and suggestions. We thank Ronald de Wolf for helpful comments. We wish to thank Nathaniel Johnston for helpful suggestions, as well as for the use of his QETLAB quantum entanglement MATLAB package \cite{qetlab}. Thanks are also due to Alessandro Cosentino, Gus Gutoski, and Christopher Perry for insightful discussions. This work was partially supported by Canada's NSERC, the US ARO, the ERC Consolidator Grant QPROGRESS, the Mike and Ophelia Lazaridis Fellowship program, and the David R. Chariton Graduate Scholarship program.

%%%%%%%%%%%%%%%%%%%%
%%
%% Bibliography
%%

\bibliographystyle{alpha}
\bibliography{refs}

\appendix

\section{Mathematical derivations}

%-----------------------------------------------------------------------------%
\subsection{Verification of procedure to group the starting state and the final measurement into a single variable}
\label{app:proof-lemma-mw12}
%-----------------------------------------------------------------------------%
Consider first the case where we have a matrix $A \in \Lin(\X \otimes \Z)$ that corresponds to a rank-1 operator that transforms a state of the computational basis into another one. Let it be equal to $\ketbra{a}{c} \otimes \ketbra{b}{d}$, with $\ketbra{a}{c} \in \Lin(\X)$,  $\ketbra{b}{d} \in \Lin(\Z)$. The channel $\Psi_{A} : \Lin(\Z) \rightarrow \Lin(\X)$ such that $J(\Psi_{A}) = \overline{A}$ is then the one that maps  $\ketbra{b}{d} \in \Lin(\Z)$ to $\ketbra{a}{c} \in \Lin(\X)$, and everything else in the computational basis for $\Lin(\Z)$ to 0.

\noindent Consider now an operator $M \in \lin{\Y \otimes \Z}$, and a channel $\Phi : \Lin(\X) \rightarrow \Lin(\Y)$. We want to verify that 
\begin{align}
\ip{M}{ \left( \Phi \otimes \I \right) (A)}  = \ip{\left( \I \otimes \Psi_{A} \right)(M) }{J(\Phi)}. \label{eq:formula-lemma-mw-12}
\end{align}

\noindent To do so, consider a computational basis decomposition $M=\sum_{i, j, k, l} m_{i,j,k,l}  \ketbra{i}{j}  \otimes \ketbra{k}{l}$, with $\ketbra{i}{j} \in \lin{\Y}$,  $\ketbra{k}{l} \in \lin{\Z}$. Then, the left hand side of \eqref{eq:formula-lemma-mw-12} is equal to

\begin{align}
\ip{\sum_{i, j, k, l} m_{i,j,k,l}  \ketbra{i}{j}  \otimes \ketbra{k}{l}}{\Phi(\ketbra{a}{c}) \otimes \ketbra{b}{d}} = \ip{\sum_{i, j} m_{i,j,b,d}  \ketbra{i}{j}}{\Phi(\ketbra{a}{c})} \nonumber ,
\end{align}

\noindent and the right hand side of  \eqref{eq:formula-lemma-mw-12} is equal to 

\begin{align}
 \ip{ \( \I \otimes \Psi_{A} \) \(\sum_{i, j, k, l} m_{i,j,k,l}  \ketbra{i}{j}  \otimes \ketbra{k}{l}\) }{J(\Phi)} & = \ip{\sum_{i, j} m_{i,j,b,d}  \ketbra{i}{j}  \otimes \ketbra{a}{c} }{J(\Phi)} \nonumber \\
& = \ip{\sum_{i, j} m_{i,j,b,d}  \ketbra{i}{j}}{\Phi(\ketbra{a}{c})}  \nonumber ,
\end{align}

\noindent so \eqref{eq:formula-lemma-mw-12} holds.

\eqref{eq:formula-lemma-mw-12} does extend by linearity to any choice of $A \in \Lin(\X \otimes \Z)$. Indeed, assume that it holds for $A, B  \in \Lin(\X \otimes \Z)$, and consider a linear combination $\lambda_A A + \lambda_B B$, with $\lambda_A, \lambda_B \in \mathbb{C}$. Then, the left hand side of \eqref{eq:formula-lemma-mw-12} will be given by 

\begin{align}
		\ip{M}{ \left( \Phi \otimes \I \right) (\lambda_A A + \lambda_B B)}  & =  \lambda_A \ip{M}{ \left( \Phi \otimes \I \right) (A)} +    \lambda_B \ip{M}{ \left( \Phi \otimes \I \right) (B)} \nonumber \\
		& =  \lambda_A \ip{\left( \I \otimes \Psi_{A} \right)(M) }{J(\Phi)} +  \lambda_B \ip{\left( \I \otimes \Psi_{B} \right)(M) }{J(\Phi)} \nonumber \\
		& =\ip{ \overline{\lambda_A} \left( \I \otimes \Psi_{A} \right)(M) + \overline{\lambda_B}\left( \I \otimes \Psi_{B} \right)(M) }{J(\Phi)} \nonumber .
\end{align}

We want to prove then that 

\[
\overline{\lambda_A} \left( \I \otimes \Psi_{A} \right)(M) +\overline{\lambda_B}\left( \I \otimes \Psi_{B} \right)(M) = \left( \I \otimes \Psi_{\lambda_A A + \lambda_B B} \right)(M). \]

To do so, we use the third property of the Choi representation introduced in Lemma~1, and express $ \overline{\lambda_A} \left( \I \otimes \Psi_{A} \right)(M) + \overline{\lambda_B} \left( \I \otimes \Psi_{B} \right)(M)$ as

\begin{align}
& \overline{\lambda_A}  \tr_{\Y \otimes \Z} \( J(\I_{\Lin(Y)} \otimes \Psi_{A}) (\I_{\X \otimes \Z} \otimes M^T) \) +  \nonumber \\
 & \overline{\lambda_B}  \tr_{\Y \otimes \Z} \( J(\I_{\Lin(Y)}\otimes \Psi_{B}) (\I_{\X \otimes \Z} \otimes M^T) \)    \nonumber \\
= & \tr_{\Y \otimes \Z} \( \( \overline{\lambda_A}  J(\I_{\Lin(Y)}\sigma \otimes \Psi_{A})  +\overline{\lambda_B}   J(\I_{\Lin(Y)} \otimes \Psi_{B}) \) (\I_{\X \otimes \Z} \otimes M^T) \)    \nonumber \\
= & \tr_{\Y \otimes \Z} \( \(  J(\I_{\Lin(Y)}) \otimes \overline{\lambda_A} \overline{A}  +  J(\I_{\Lin(Y)})\otimes \overline{\lambda_B} \overline{B} \) (\I_{\X \otimes \Z} \otimes M^T) \)   \nonumber \\
= & \tr_{\Y \otimes \Z} \( \(  J(\I_{\Lin(Y)}) \otimes \( \overline{\lambda_A} \overline{A} + \overline{\lambda_B} \overline{B} \)   \) (\I_{\X \otimes \Z} \otimes M^T) \)    \nonumber \\
= & \left( \I \otimes \Psi_{\lambda_A A + \lambda_B B} \right)(M)    . \nonumber
\end{align}

%-----------------------------------------------------------------------------%
\subsection{Derivation for Lemma~4}
\label{app:proof-lemma-strat}
%-----------------------------------------------------------------------------%

% Proof for Lemma~for 1/n strategy.

\begin{proof}
Given that $n$ parallel repetitions of the game are considered, our claim states that Bob will win {\it at least} one out of the $n$ repetitions if he adopts $\Lambda_{n}$  as his strategy when the projective measurement made by Alice corresponds to the parameter $\theta_{n,\alpha}$. A similar argument also holds for $\Xi_{n}$ at the corresponding angle $\gamma_{n,\alpha} $. We prove this explicitly for the strategy $\Lambda_{n}$ , and the other case follows using the same argument. The proof of this lemma uses a technique of conditioning where we consider the resulting state conditioned on Bob obtaining a losing outcome in the first projective measurement of Alice, and the corresponding probability for such an outcome. Then, we generalize this procedure to the rest of the parallel repetitions. To conclude the proof, we set the probability of the ``all-losing state" at the end to zero, which allows us to solve for $\theta$ in the final equation.

 First, let us define the pure states:

\begin{equation}
\label{eq:losingvectorsstrategy}
	\begin{aligned}
		v_{\theta}= \cos(\theta) \ket{00} + \sin(\theta) \ket{11}, &\quad s_{\theta} = \ket{01}, \\
		\quad w_{\theta} = \sin(\theta)\ket{00} - \cos(\theta) \ket{11}, &\quad t_{\theta}  = \ket{10},
	\end{aligned}
\end{equation}

\noindent where we recall from Section \ref{sec:general-model}  that $v_{\theta} \in \Y \otimes \Z$ is the state which corresponds to the winning projective measurement outcome, and $w_{\theta}, s_{\theta}$, and $t_{\theta} \in \Y \otimes \Z$ are the states that correspond to the losing projective measurement. Essentially, Bob is trying then to transform the state prepared by Alice to something as close as possible to $v_{\theta}$, while restricted to operating on one half on the state.

Let $\Lambda_n$ be the operator defined as
\begin{align}
	\Lambda_n = \sum_{r \in \{0,1\}^n} \left(-1\right)^{\land r + \oplus r} \ketbra{r}{r}, 
\end{align}
$\Lambda_n'$ be the similar operator 
\begin{align}
	\Lambda_n^{\prime} = \sum_{r \in \{0,1\}^n} \left(-1\right)^{\oplus r} \ketbra{r}{r}.
\end{align}
and define the vector $\kappa_n$ as
\begin{align}
	\kappa_n = \sum_{a \in \{0,1\}^n} \bigotimes_{i=0}^{n-1} \alpha^{(1-a_i)} \left(1-\alpha^2\right)^{a_i/2} \ket{a_i a_i}.
\end{align}
We run now through the parallel repetition of $n$ copies of the game. Since the initial shared state is $u^{\otimes n}_{\alpha} = \(\alpha \ket{00} + \sqrt{1 - \alpha^2} \ket{11}\)^{\otimes n}$, the state after Bob applies his channel (acting on his qubits for all of the $n$ parallel repetitions) is
\begin{align}
	f_{\alpha}^0 = \left( \Lambda_n \otimes \I_{\Z_1 \otimes \cdots \otimes \Z_n} \right) \kappa_n
\end{align}

\noindent We shall condition now on Bob losing the first out of $n$ parallel repetitions. It should be noted that since Alice starts with the entangled state $u^{\otimes n}_{\alpha}$ and Bob performs a unitary diagonal operation, the states $s_{\theta}$ and $t_{\theta}$ in \eqref{eq:losingvectorsstrategy} do not contribute to the losing projective measurement outcome. Once we condition on Bob losing the first game, the resulting state is then a normalization of	
\begin{align}
	f_{\alpha,\theta}^1 & =  \left( w_{\theta} w_{\theta}^* \otimes \I \right) f_{\alpha}^0  \nonumber \\
	&=  w_{\theta} \otimes \alpha \sin(\theta) \left(\Lambda^{\prime}_{n-1} \otimes \I_{\Z_2 \otimes \cdots \otimes \Z_n} \right) \kappa_{n-1}  \nonumber \\
	& ~~ + w_{\theta} \otimes \sqrt{1-\alpha^2} \cos(\theta)\left( \Lambda_{n-1} \otimes \I_{\Z_2 \otimes \cdots \otimes \Z_n} \right) \kappa_{n-1},\label{eq:precondition1}
\end{align}	

\noindent with the associated probability being $(f_{\alpha, \theta}^1)^*f_{\alpha, \theta}^1$.

Generalizing this to Bob losing all $n$ games, one can observe that the $-1$'s for the $\cos(\theta)$ term in $w_{\theta}$ cancel the negative terms from the $(-1)^{\bigoplus r}$ term in $\Lambda_n$, and Equation~\eqref{eq:precondition1} generalizes to:

\begin{align}
f_{\alpha, \theta}^n&= (w_{\theta})^{\otimes n} \Big( \alpha^n\sin(\theta)^n+n(\alpha^{n-1}\sqrt{1-\alpha^2} )\sin(\theta)^{n-1}\cos(\theta)+\ldots \nonumber \\
 & \quad + n(\alpha(1-\alpha^2)^{(n-1)/2} )\cos(\theta)^{n-1}\sin(\theta)- (1-\alpha^2)^{n/2}\cos(\theta)^n\Big)\\
 & = (w_{\theta})^{\otimes n} \((\alpha\sin(\theta)+\sqrt{1-\alpha^2}\cos(\theta))^n-2(1-\alpha^2)^{n/2}\cos(\theta)^n\).
\end{align}

In order for Bob to ensure he  wins {\it at least} 1 out of the $n$ games  with certainty, we require that $\left \lVert f_{\alpha, \theta}^n \right \rVert = 0$, which~implies:

\begin{align}
(\alpha\sin(\theta)+\sqrt{1-\alpha^2}\cos(\theta))^n-2(1-\alpha^2)^{n/2}\cos(\theta)^n&=0.
\end{align}

This implies that for the angle $\theta_{n, \alpha}=\tan^{-1}\Big(\sqrt{\frac{1}{\alpha^2}-1}\(2^{1/n}-1\)\Big)$, the strategy corresponding to $\Lambda_{n}$ gives us a perfect hedging strategy. Following the same procedure, using the strategy corresponding to $\Xi_{n}$  yields the similar condition that:
\begin{equation}
\begin{aligned}
(\alpha\sin(\theta)+\sqrt{1-\alpha^2}\cos(\theta))^n-2\alpha^{n}\sin(\theta)^n&=0,
\end{aligned}
\end{equation}
giving us as a solution $\gamma_{n, \alpha} =  \tan^{-1} \left( \sqrt{\frac{1}{\alpha^2} - 1} \left( \frac{1}{2^{1/n} - 1} \right) \right)$.
\end{proof}

%-----------------------------------------------------------------------------%
\subsection{Derivation for Lemma 5}
\label{app:proof-lemma-unitary}
%-----------------------------------------------------------------------------%

\begin{proof}
As in the previous proof, to win at least 1 out of $n$ games, Bob needs to avoid the outcome corresponding to the state $(\sin(\theta)\ket{00} - \cos(\theta)\ket{11})^{\otimes n}$ (other states for the losing outcome can be ignored since Bob's strategy corresponds to a diagonal matrix). Let us now define a matrix 

\begin{equation}
\begin{aligned}
D=\sum_{r \in \{0,1\}^n}  (-1)^{\abs{r}}\sin(\theta)^{n-\abs{r}} \cos(\theta)^{\abs{r}}  \ketbra{r}{r},
\end{aligned}
\end{equation}

\noindent such that $(\sin(\theta)\ket{00} - \cos(\theta)\ket{11})^{\otimes n} =\vecnotation{D}$. For convenience, we denote $\lambda=\tan(\theta)$, and rewrite $D$ as
\begin{equation}
\begin{aligned}
D = \cos(\theta)^n\sum_{r \in \{0,1\}^n} (-1)^{\abs{r}}\lambda^{n-\abs{r}}  \ketbra{r}{r}.
\end{aligned}
\end{equation}

\noindent We also introduce an operator

\begin{equation}
\begin{aligned}
F=\sum_{r \in \{0,1\}^n} (1-\alpha^2)^{\abs{r}/2}\alpha^{n-\abs{r}}  \ketbra{r}{r},
\end{aligned}
\end{equation}

\noindent such that $u^{\otimes n}_{\alpha} = \vecnotation{F}$, where $u_{\alpha}$ is again the pure state $\alpha \ket{00} + \sqrt{1 - \alpha^2} \ket{11}$ shared by Alice and Bob at the beginning of a single repetition of the protocol.

From our construction the unitary $U$ that Bob applies in Lemma~5 to his portion of the entangled state $u^{\otimes n}_{\alpha}$ is
\begin{equation}
\begin{aligned}
U=(-1)^n |0\rangle \langle 0| - |1\rangle \langle 1 |  +\sum_{i=1}^{n-1} \sum_{\substack{r \in \{0,1\}^{n} \\ \abs{r}=i}} (-1)^{n+i}k_{r}\ket{r}\bra{r}.
\end{aligned}
\end{equation}

The state that Alice holds before measurement is then $(U \otimes \I_{\Z_{1 \ldots n}})u^{\otimes n}_{\alpha}$. We analyze how successful the application of this channel would be to avoid $(\sin(\theta)\ket{00} - \cos(\theta)\ket{11})^{\otimes n}$. Upon explicit computation of the formula $\langle \vecnotation{D}, (U\otimes \I_{\Z_{1 \ldots n}})\vecnotation{F}\rangle $, and using repeatedly the fact that $\vecnotation{V}=(V\otimes \I)\vecnotation{\I}$, we obtain $\ip{\vecnotation{D}}{\vecnotation{UF}}$, which is equal to $\ip{D}{UF}$ by the properties of the $\setft{vec}$ operator, resulting in the following expression:

\begin{align}
	\ip{D}{UF}&=\tr\Big( (-1)^n \alpha^n\lambda^n |0^n \rangle \langle 0^n| + (1-\alpha^2)^{n/2} (-1)^{n+1}|1^n\rangle \langle 1^n| \nonumber \\
	  & ~~+ \sum_{i=1}^{n-1} \sum_{\substack{r \in \{0,1\}^{n} \\ \abs{r}=i}}(-1)^{n}k_{r}  (1-\alpha^2)^{i/2} \alpha^{n-i} \lambda^{n-i} \ket{r}\bra{r} \Big) \nonumber \\
&=(-1)^n  \alpha^n \tr \Big( \lambda^n |0^n\rangle \langle 0^n| - \left (\sqrt{\frac{1}{\alpha^2}-1}\right )^n |1^n\rangle \langle 1^n|  \nonumber \\
&~~+ \sum_{i=1}^{n-1} \sum_{\substack{r \in \{0,1\}^{n} \\ \abs{r}=i}} k_{r} \left (\sqrt{\frac{1}{\alpha^2}-1}\right )^i \lambda^{n-i} \ket{r}\bra{r} \Big)  \nonumber \\
&=(-1)^n  \alpha^n \( \lambda^n  - \left (\sqrt{\frac{1}{\alpha^2}-1}\right )^n + \sum_{i=1}^{n-1} \sum_{\substack{r \in \{0,1\}^{n} \\ \abs{r}=i}}k_{r} \left (\sqrt{\frac{1}{\alpha^2}-1}\right )^i \lambda^{n-i}  \) \nonumber \\
&=(-1)^n  \alpha^n \left( \sqrt{\frac{1}{\alpha^2}-1}\right)^n  \( \lambda_{\alpha}^n  -  1 + \sum_{i=1}^{n-1} \sum_{\substack{r \in \{0,1\}^{n} \\ \abs{r}=i}}k_{r} \lambda_{\alpha}^{n-i}  \) \label{eq:LambdaEquation},
\end{align}

\noindent where $\lambda_{\alpha} = \lambda \cdot \left( \sqrt{\dfrac{1}{\alpha^2}-1}\right)^{-1}$.

Note that for the range of $\theta$ we are considering, it holds that $  2^{1/n} - 1 \leq  \lambda_{\alpha}  \leq   \dfrac{1}{2^{1/n} -1}$. Note as well that from our choice of $k_{r}$, for all $i$ we have that $\operatorname{Im}\(\sum_{\substack{r \in \{0,1\}^{n} \\ \abs{r}=i}}k_{r} \lambda_{\alpha}^{n-i} \)=0$, and therefore the imaginary part of $~\eqref{eq:LambdaEquation}$ is equal to 0. It then suffices to prove that for any choice of $\lambda_a$ and $n$, there exists an $s_{\theta, \alpha, n} \in [-1,1]$ such that, when plugged into the definition of 
$ k_{r}$  in the statement of Lemma~5  we have 
\begin{align}
\label{eq:LambdaEquation2}
\lambda_{\alpha}^n  -  1 + \sum_{i=1}^{n-1} \sum_{\substack{r \in \{0,1\}^{n} \\ \abs{r}=i}}\operatorname{Re} \(k_{r}\) \lambda_{\alpha}^{n-i} = 0.
\end{align}

Now, as the left hand side of~\eqref{eq:LambdaEquation2} is an affine function of $s_{\theta, \alpha, n}$ with a positive linear coefficient, to prove the existence of such an $s_{\theta, \alpha, n}$, it suffices to prove that the left hand side of~\eqref{eq:LambdaEquation2}) $\leq 0$ when $s_{\theta, \alpha, n}=-1$ , and that the left hand side of~\eqref{eq:LambdaEquation2} $\geq 0$ when $s_{\theta, \alpha, n}=1$.

We look first into the case when $s=-1$. Then, when $1 \leq \lambda_{\alpha} \leq  \dfrac{1}{2^{1/n} -1}$ it holds that:

\begin{align}
\lambda_{\alpha}^n  -  1 + \sum_{i=1}^{n-1} \sum_{\substack{r \in \{0,1\}^{n} \\ \abs{r}=i}}\operatorname{Re} \(k_{r}\) \lambda_{\alpha}^{n-i} & = \lambda_{\alpha}^n  -  1 - \sum_{i=1}^{n-1} {n \choose {n-i}} \lambda_{\alpha}^{n-i} \nonumber \\
& =  2\lambda_{\alpha}^n - \lambda_{\alpha}^n  - 1 - \sum_{i=1}^{n-1}  {n \choose {n-i}} \lambda_{\alpha}^{n-i} \nonumber \\
& =  2\lambda_{\alpha}^n - (1+\lambda_{\alpha})^n,
\end{align}

\noindent which is $\leq 0$ whenever $\lambda_{\alpha} \leq   \dfrac{1}{2^{1/n} -1}$. When $2^{1/n} - 1 \leq \lambda_{\alpha} < 1$,  that the left hans side of~\eqref{eq:LambdaEquation2} $\leq 0$ follows from two simple facts. First, the fact that $\lambda_{\alpha}^n < 1$, so $\lambda_{\alpha}^n - 1 < 0$ . Second, the fact that for each $ \sum_{\substack{r \in \{0,1\}^{n} \\ \abs{r}=i}}\operatorname{Re} \(k_{r}\) \lambda_{\alpha}^{n-i} $ term, $ \sum_{\substack{r \in \{0,1\}^{n} \\ \abs{r}=i}}\operatorname{Re} \(k_{r}\) \leq  - {n \choose i} + 1 \leq 0$.

We look now into the case when $s=1$. Then, when $2^{1/n} -1 \leq \lambda_{\alpha} < 1$ it holds that:

\begin{align}
\lambda_{\alpha}^n  -  1 + \sum_{i=1}^{n-1} \sum_{\substack{r \in \{0,1\}^{n} \\ \abs{r}=i}}\operatorname{Re} \(k_{r}\) \lambda_{\alpha}^{n-i} & = \lambda_{\alpha}^n  -  1 + \sum_{i=1}^{n-1} {n \choose {n-i}} \lambda_{\alpha}^{n-i} \nonumber \\
& =  -2 + \lambda_{\alpha}^n  + 1 + \sum_{i=1}^{n-1}  {n \choose {n-i}} \lambda_{\alpha}^{n-i} \nonumber \\
& =  -2 + (1+\lambda_{\alpha})^n,
\end{align}

\noindent which is $\geq 0$ whenever $\lambda_{\alpha} \geq 2^{1/n} -1$. When $1 \leq \lambda_{\alpha} \leq  \dfrac{1}{2^{1/n} -1}$,  that the left hand side of \eqref{eq:LambdaEquation2} $\geq 0$ follows from two simple facts. First, the fact that $\lambda_{\alpha}^n \geq 1$. Second, the fact that for each $ \sum_{\substack{r \in \{0,1\}^{n} \\ \abs{r}=i}}\operatorname{Re} (k_{r}) \lambda_{\alpha}^{n-i} $ term, it is the case that $ \sum_{\substack{r \in \{0,1\}^{n} \\ \abs{r}=i}}\operatorname{Re} (k_{r}) \geq {n \choose i} - 1$.

\end{proof}

%-----------------------------------------------------------------------------%
\subsection{Derivation for Lemma~6}
\label{app:proof-theorem-outside}
%-----------------------------------------------------------------------------%

\begin{proof}

We will consider here the case where $\theta < \theta_{n, \alpha}$. The other case proceeds similarly. 

Remember first that we characterized the chance of achieve $1$-out-of-$n$ hedging by the following SDP program in Section \ref{sec:general-model}:

%\noindent\begin{minipage}{\textwidth}
%\begin{minipage}[c][6cm][c]{\dimexpr0.5\textwidth-5pt\relax}

%\vspace{-3cm}
\begin{center}
    \centerline{\underline{$m_{n,\alpha,\theta}$: Primal problem}}\vspace{-5mm}
    \begin{equation}
   \label{eq:1OutOfNSDPApl}
    \begin{aligned}
      \text{minimize:}\quad & \ip{Q_{0,\alpha,\theta}^{\otimes n}}{X} \\
      \text{subject to:}\quad & \tr_{\Y_1 \otimes \cdots \otimes \Y_n}(X) = \I_{\X_1 \otimes \cdots \otimes \X_n},\\
      & X \in \Pos(X_1 \otimes \Y_1 \otimes \cdots \otimes \X_n \otimes \Y_n).
    \end{aligned}
    \end{equation}
\end{center}

%\end{minipage}\hfill
%\begin{minipage}[c][2cm][c]{\dimexpr0.5\textwidth-5pt\relax}

\newpage
%\vspace{-3cm}
\begin{center}
    \centerline{\underline{$m_{n,\alpha,\theta}$: Dual problem}}\vspace{-5mm}
    \begin{equation}
   \label{eq:1OutOfNSDPDualApl}
    \begin{aligned}
      \text{maximize:}\quad & \tr(Y) \\
      \text{subject to:}\quad & \pi \left(\I_{\Y_1 \otimes \cdots \otimes \Y_n} \otimes Y\right)\pi^{*} \leq Q_{0,\alpha,\theta}^{\otimes n},\\
      & Y \in \Herm(\X_1 \otimes \cdots \otimes \X_n).
    \end{aligned}
    \end{equation}
\end{center}
%\end{minipage}%
%\end{minipage}

%\vspace{-2.5cm}

Then, to prove that perfect hedging is not possible when $\theta < \theta_{n, \alpha}$, we prove the feasibility  in the dual SDP ~\eqref{eq:1OutOfNSDPDuaApl} of an operator $Y$ with positive objective value. This operator is obtained from applying complementary slackness conditions to the primal solution corresponding to $\Lambda_n$. Therefore, it has value for the dual equal to the value in the primal SDP~\eqref{eq:1OutOfNSDPApl} for the solution corresponding to $\Lambda_n$. By weak duality, its feasibility proves then the optimality of $\Lambda_n$ when $\theta < \theta_{n, \alpha}$.

To prove the feasibility of $Y$, we will express $Q_{0,\alpha,\theta}^{\otimes n} - \pi \left(\I_{\Y_1 \otimes \cdots \otimes \Y_n} \otimes Y\right)\pi^{*}$ as a direct sum of smaller matrices. This reduces the question about feasibility of $Y$ to a question about the positive-semidefiniteness of these smaller matrices. Each of these smaller matrices will have all proper leading principal minors be positive semi-definite, so by Sylvester's criterion it will suffice to check that their determinant is non-negative. We will then obtain a closed formula for these determinants, and prove that they are indeed non-negative.

We will first consider the case with $\alpha = 1/\sqrt{2}$, and then give an overview of the small changes involved in adapting the proof to other values of $\alpha$. To simplify our argument, we will incur in a bit of notation abuse in this section, and omit the permutation operators in the definition of the dual SDP~\eqref{eq:1OutOfNSDPDualApl} that remind us that matrices at the sides of a $\leq$ inequality must have their entries reordered to make the spaces on which they are defined be in the same order at both sides of the inequality.

\subsubsection{Study of $Q_{0,1/\sqrt{2},\theta}^{\otimes n} $}

$Q_{0, \alpha, \theta} \in \Pos(\X \otimes \Y)$ is given by $\ket{\psi_0^1}\bra{\psi_0^1}+\ket{\psi_0^2}\bra{\psi_0^2}+\ket{\psi_0^3}\bra{\psi_0^3}$, where the $\ket{\psi_0^i}$ are defined as
\begin{equation}
	\begin{aligned}
		 \ket{\psi_0^1} &=  \alpha \sin(\theta)\ket{00} - \sqrt{1-\alpha^2}\cos(\theta) \ket{11}, \\
		\ket{\psi_0^2} & =  \alpha \ket{01}, \\
		 \ket{\psi_0^3} & =  \sqrt{1-\alpha^2}\ket{10}.
	\end{aligned}
\end{equation}

\noindent This follows from considering the definition of $P_{0, \theta}$ given in Section \ref{sec:general-model}, and observing that the operator $\Psi_{\rho_{\alpha}}$ satisfying $J(\Psi_{\rho_{\alpha}}) = \overline{u_{\alpha}u_{\alpha}^*}$ (with $u_{\alpha}=\alpha \ket{00} + \sqrt{1 - \alpha^2} \ket{11}$  the initial state shared between Alice and Bob) maps a state $\sigma \in \density{\Z}$ to $( \alpha \ketbra{0}{0} + \sqrt{1-\alpha^2}\ketbra{1}{1}) \sigma ( \alpha \ketbra{0}{0} + \sqrt{1-\alpha^2}\ketbra{1}{1})$. We can then write $Q_{0,1/\sqrt{2},\theta}^{\otimes n}$ as

\begin{align}
		 Q_{0,1/\sqrt{2},\theta}^{\otimes n} & =&&  \(\dfrac{1}{2}\)^{n} \Big( ( \sin(\theta)\ket{00} - \cos(\theta)\ket{11})  (\sin(\theta)\bra{00} -  \cos(\theta) \bra{11}) \nonumber \\ 
		 &  &&+ \ketbra{01}{01} + \ketbra{10}{10} \Big)^{\otimes n}   \label{q0n:firstSum} ~	\\
		& = &&  \(\dfrac{1}{2}\)^n \sum_{a, b, c, d \in \{0,1\}^{n}} \ket{a}\ket{b}\bra{c}\bra{d} \prod_{i=0}^{n-1} \Big( \delta_{c_i, 1-d_i}\delta_{a_i, c_i}\delta_{b_i, d_i} \nonumber \\
		& & & +  \delta_{a_i,b_i}\delta_{c_i, d_i} \Big(
	 \delta_{a_i,1-c_i} (-\sin(\theta)\cos(\theta)) + \delta_{a_i,c_i}\delta_{a_i,1} \cos(\theta)^2  \nonumber \\  & & & ~~ +  \delta_{a_i,c_i}\delta_{a_i,0} \sin(\theta)^2   \Big) \Big) \nonumber  \\		
		& = &&  \label{q0n:secondSum}  \(\dfrac{1}{2}\)^{n} \sum_{a, c \in \{0,1\}^{n}} \ketbra{a}{c} \otimes \sum_{b, d \in \{0,1\}^{n}}  \ketbra{b}{d} \prod_{i=0}^{n-1}   \Big( \delta_{a_i, 1-b_i}\delta_{c_i, 1- d_i}\delta_{a_i, c_i} + \nonumber \\ 		
& &&  \delta_{a_i,b_i}\delta_{c_i, d_i} \Big( \delta_{a_i,1-c_i} (-\sin(\theta)\cos(\theta)) +    \delta_{a_i,c_i}\delta_{a_i,1} \cos(\theta)^2  \nonumber \\
& & & ~~ + \delta_{a_i,c_i}\delta_{a_i,0} \sin(\theta)^2   \Big) \Big)  .
\end{align}

The key insight to go ahead with the proof is to notice that this matrix can be written as a direct sum of $3^n$ smaller matrices. Indeed, observe that \eqref{q0n:firstSum} can be equivalently written as

\begin{align}
\dfrac{1}{2^n}  \sum_{w \in \{0, 1,2\}^n} \bigotimes_{i=0}^{n-1} \ketbra{\psi_{w_i}}{\psi_{w_i}}, \text{ where } \ket{\psi_{w_i}} = \begin{cases}
							\sin(\theta)\ket{00} - \cos(\theta)\ket{11}, \text{ if } w_i=0 \\
							\ket{01}, \text{ if } w_i=1 \\
							\ket{10}, \text{ if } w_i=2 \\
							\end{cases} \label{q0n:thirdSum} .
\end{align}

Then, the coefficient for each $\ketbra{a}{c} \otimes \ketbra{b}{d} $ term in the summation in \eqref{q0n:secondSum} will receive contribution from at most one of the elements in \eqref{q0n:thirdSum}. This element will be the one with $w_i= \begin{cases} 0 \text{ if } a_i =b_i \\ 1 \text{ if }(a_i, b_i) = (0,1) \\ 2 \text{ if } (b_i, a_i) = (1,0) \end{cases}$.

Since this only depends on $\ket{ab}$, all elements on the same row of $Q_{0,1/\sqrt{2},\theta}^{\otimes n}$ come from the same term in \eqref{q0n:thirdSum}. As each row of $Q_{0,1/\sqrt{2},\theta}^{\otimes n} $ has at least one non-zero term,  \eqref{q0n:thirdSum} implies then a decomposition $Q_{0,1/\sqrt{2},\theta}^{\otimes n} $ into a direct sum of smaller matrices, each of them with rank 1.

We can then identify each of these matrices by the corresponding  choice of $w$ in \eqref{q0n:thirdSum}. We will do so by writing them as $Q_{0,1/\sqrt{2},\theta}^{\otimes n} (w)$. We denote the number of $0$s, $1$s and $2$s in $w$ by $n_0(w)$, $n_1(w)$ and $n_2(w)$, respectively. Also, note that there will be $3^n$ matrices in our decomposition, with the dimension of $Q_{0,1/\sqrt{2},\theta}^{\otimes n} (w)$ being given by $2^{n_0(w)}$. Also, note that the number of matrices of size $2^k$ is given by ${n \choose k}2^{n-k}$. This corresponds to choosing on which $k$ positions $w_i=0$, and what is the value of $w_i$ for the other ones.

It will be convenient later to have a formula for the restriction to the diagonal of $Q_{0,1/\sqrt{2},\theta}^{\otimes n} (w)$. Using the description in \eqref{q0n:thirdSum}, we have that it is given by

\begin{align}
	\(\dfrac{1}{2}\)^{n} \sum_{w' \in M_w \subseteq \{0, 1\}^n}  g(w,w') \ket{w'}\ket{f(w,w')}\bra{w'}\bra{f(w,w')} \label{directSum:restrictionToDiagonal}
\end{align}

\noindent where $M_w$ is given by the cartesian product $\bigtimes_{i=0}^{n-1} M_{w_i}$,
with $\begin{cases} 
M_0 = \{0, 1\} \\
M_1 = \{0\} \\
M_2 = \{1\}  \\
\end{cases}$,\\
$g(w,w')~=~\prod_{i=0}^{n-1} g(w_i, w_i')$~with~$\begin{cases}
 g(0, 0) = \sin^2(\theta) \\
 g(0, 1) = \cos^2(\theta) \\
 g(1, 0) = 1 \\
 g(2, 1) = 1 \\
 \end{cases}$,  $f(w,w')_i = \begin{cases}
w'_i \text{ if }  w_i = 0 \\
1 - w'_i \text{ if } w_i = 1
\end{cases}$. \\
\noindent Note that by definition of $M_w$, it is not necessary to define $g(w_i, w_i')$ for values of $(w_i, w_i')$ not included in our definition of g.

\subsubsection{Study of our candidate for $Y$ in the $\alpha=1/\sqrt{2}$ case}

We define now our candidate solution $Y$ for the dual problem, given by

\begin{align}
Y= -\epsilon\( \( \dfrac{1}{\sqrt{2}}\sin(\theta)\ket{0}\bra{0} + \dfrac{1}{\sqrt{2}}\cos(\theta)\ket{1}\bra{1}\)^{\otimes n} - 2 \(\dfrac{1}{\sqrt{2}}\cos(\theta) \ket{1}\bra{1}\)^{\otimes n} \),
\end{align}

\noindent where $\epsilon$ is a value $ > 0$  given by $\(\dfrac{1}{2}\)^{n/2}\( 2\cos(\theta)^n - (\cos(\theta) + \sin(\theta))^n \)$. Note that the definition of $\theta_{n,1/\sqrt{2}}$ implies that this value is positive indeed for $\theta < \theta_{n,1/\sqrt{2}}$. We can then write $Y$ as 

\begin{align}
\sum_{a \in \{0,1\}^{n}} \lambda_{a} \ket{a}\bra{a}, \text{where } \lambda_{a} =  \begin{cases} 
      -\epsilon \(\dfrac{1}{2}\)^{n/2} \sin(\theta)^{n-\abs{a}}\cos(\theta)^{\abs{a}} & \text{for } a \neq 1^n  \\
      \epsilon \(\dfrac{1}{2}\)^{n/2} \cos(\theta)^n & \text{for } a=1^n  \\
   \end{cases}
\label{dualSolution:lambdas}
\end{align}

\noindent Note that its trace (i.e., its value for the dual program) is given by 
\begin{equation}
\begin{aligned}
- \(\dfrac{1}{2}\)^{n/2} \epsilon \Big((\sin(\theta) + \cos(\theta))^n  - 2 \cos(\theta)^n \Big),
\end{aligned}
\end{equation}

\noindent which will again be positive for $\theta < \theta_{n,1/\sqrt{2}}$ by definition of $\theta_{n,1/\sqrt{2}}$.

This $Y$ has been obtained from the strategy $\Lambda_n$ in Lemma~4, and its feasibility proves the optimality of $\Lambda_n$ for $\theta < \theta_{n, 1/\sqrt{2}}$. This is an example of complementary slackness behavior, and follows from an observation \cite{cs766solutions} that given a feasible solution $X$ to the primal SDP~\eqref{eq:1OutOfNSDPApl}, $\tr_{\Y_1 \otimes \cdots \otimes \Y_n}(Q_{0,\alpha,\theta}^{\otimes n}X)$ is an operator with the same objective value for the dual SDP~\eqref{eq:1OutOfNSDPDualApl}. Furthermore,  $\tr_{\Y_1 \otimes \cdots \otimes \Y_n}(Q_{0,\alpha,\theta}^{\otimes n}X)$ satisfies the feasibility constraints of the dual if and only if $X$ represents an optimal solution to the primal. Therefore, after we experimentally observed that $\Lambda_n$ seemed to be optimal for $\theta < \theta_{n, \alpha}$ to obtain our proposed $Y$ we computed the corresponding value of  $\tr_{\Y_1 \otimes \cdots \otimes \Y_n}(Q_{0,1/\sqrt{2},\theta}^{\otimes n}X)$.  $X$ is given in this computation by the primal solution that represents the channel for the unitary in $\Lambda_n$,
\begin{align}
X= \sum_{i,j \in \{0,1\}^n} \ket{ii} \bra{jj}  (-1)^{\wedge i+\bigoplus i + \wedge j +\bigoplus j}.
\end{align}

\subsubsection{Feasibility of $Y$ in the $\alpha=1/\sqrt{2}$ case}

We want to prove that $Y$ is feasible - that is to say, $Q_{0,1/\sqrt{2},\theta}^{\otimes n}  - Y \otimes \I\geq 0$. Since $Y$ is diagonal, the direct sum decomposition of $Q_{0,1/\sqrt{2},\theta}^{\otimes n}$ corresponds to a direct sum decomposition of $Y$.  Since positive semidefiniteness is preserved by the direct sum operator, it is then enough to prove that each of the $S_w=  Q_{0,1/\sqrt{2},\theta}^{\otimes n}(w) - (Y \otimes \I)(w)$ matrices are positive semidefinite, where $(Y \otimes \I)(w)$ denotes $Y \otimes \I$ restricted to the rows/columns of $Q_{0,1/\sqrt{2},\theta}^{\otimes n}$ assigned to $Q_{0,1/\sqrt{2},\theta}^{\otimes n}(w)$.

Consider first the largest of these matrices. This will be $S_{0^n}$, with size $2^n$. Using \eqref{q0n:thirdSum}, we have that it is given by
\begin{equation}
\begin{aligned}
S_{0^n} =   \(\dfrac{1}{2}\)^{n} \sum_{a, c \in \{0,1\}^{n}} \ket{aa} \bra{cc} \Big( \prod_{i=0}^{n-1}   \Big(\delta_{a_i,1-c_i} \cdot -\sin(\theta)\cos(\theta)  +  \nonumber \\
\delta_{a_i,c_i}\delta_{a_i,1} \cos(\theta)^2 +  \delta_{a_i,c_i}\delta_{a_i,0} \sin(\theta)^2   \Big)  - 2^n \lambda_a \Big).
\end{aligned}
\end{equation}

For example, for $n=2$, $S_{00}$ is given by
 \begin{align}
 	\dfrac{1}{4}
	  \begin{pmatrix}
	  \sin(\theta)^4  - 4\lambda_{00} & -\sin(\theta)^3\cos(\theta) & -\sin(\theta)^3\cos(\theta)  &  \sin(\theta)^2\cos(\theta)^2 \\ 
	  -\sin(\theta)^3\cos(\theta)  & \sin(\theta)^2\cos(\theta)^2 - 4\lambda_{01}    &   \sin(\theta)^2\cos(\theta)^2  & -\sin(\theta)\cos(\theta)^3 \\ 
	  -\sin(\theta)^3\cos(\theta) &   \sin(\theta)^2\cos(\theta)^2  & \sin(\theta)^2\cos(\theta)^2 - 4\lambda_{10}  &  -\sin(\theta)\cos(\theta)^3 \\ 
	  \sin(\theta)^2\cos(\theta)^2  &  -\sin(\theta)\cos(\theta)^3 &  -\sin(\theta)\cos(\theta)^3 & \cos(\theta)^4 - 4\lambda_{11}  \\ 
	  \end{pmatrix} \nonumber
\end{align}

Consider now that since $Q_{0,1/\sqrt{2},\theta}^{\otimes n} \geq 0$, and for $a \neq 1^n $, $\lambda_a < 0$, the first $2^n-1$ principal minors of $S_{0^n}$ are $\geq 0$. By Sylvester's criterion, to prove that $S_{0^n} \geq 0$, it suffices then to prove that $\det(S_{0^n}) \geq 0$.  Note that $\det(S_{0^n})$ is a polynomial in $\epsilon$. This polynomial has all the coefficients below the one for $\epsilon^{2^n-1}$ equal to $0$. This is because $Q_{0,1/\sqrt{2},\theta}^{\otimes n}(0^n)$ has rank 1 - therefore, each minor of it with at least two rows will have determinant equal to zero.  Using this, and going through the determinant formula, we see that $\det(S_{0^n}) $ is given by

\begin{align}
	 & \( \epsilon^{2^n-1} (-1)^{2^n -1}    \sum_{a \in \{0,1\}^{n}}    \(\dfrac{1}{2}\)^{n}   \cos(\theta)^{2\abs{a}} \sin(\theta)^{2(n - \abs{a})} \prod_{\substack{b \in \{0,1\}^n \\ b \neq a}} \dfrac{ \lambda_b}{\epsilon} \) \nonumber \\ + &   \(  \epsilon^{2^n}(-1)^{2^n}\prod_{a \in \{0,1\}^n} \dfrac{\lambda_a }{\epsilon} \)   \\
	= & \epsilon^{2^n-1} \( \epsilon - \sum_{a \in \{0,1\}^{n}}   \dfrac{  \(\dfrac{1}{2}\)^{n} \cos(\theta)^{2\abs{a}} \sin(\theta)^{2(n - \abs{a}})}{\lambda_a / \epsilon}\)  \prod_{a \in \{0,1\}^n} \dfrac{\lambda_a }{\epsilon}   \\
	= & \epsilon^{2^n-1} \( \epsilon + \sum_{a \in \{0,1\}^{n}}   \(\dfrac{1}{2}\)^{n/2} \cos(\theta)^{\abs{a}} \sin(\theta)^{n - \abs{a}} - 2 \(\dfrac{1}{2}\)^{n/2} \cos(\theta)^{n} \)   \prod_{a \in \{0,1\}^n} \dfrac{\lambda_a }{\epsilon}
\end{align}

\noindent Since all of the $\lambda_a/\epsilon$ except the one for $1^n$ are negative, we have that $\displaystyle \epsilon^{2^n-1} \prod_{a \in \{0,1\}^n} \dfrac{\lambda_a}{\epsilon} $ is negative whenever $\epsilon > 0$. Therefore,

\begin{align}
\det(S_{0^n,0^n}) \geq 0 & \iff\\
 \epsilon + \sum_{a \in \{0,1\}^{n}}   \(\dfrac{1}{2}\)^{n/2} \cos(\theta)^{\abs{a}} \sin(\theta)^{n - \abs{a}} - 2 \(\dfrac{1}{2}\)^{n/2} \cos(\theta)^{n} \leq 0& \iff \\	
\epsilon \leq  \(\dfrac{1}{2}\)^{n/2}\( 2(\cos(\theta))^n - (\cos(\theta) + \sin(\theta))^n \),
\end{align}

\noindent which is true by definition of $\epsilon$. We have then that our proposed feasible solution $Y$ produces a positive-semidefinite $S_{0^n}$. To verify the feasibility of $Y$, it remains to prove the positive-semidefiniteness of the rest of the $S_{w}$. 

To do so, consider an arbitrary $S_w$, $w \in \{0,1,2\}^n - \{0^n\}$, with a corresponding $M_w$, as defined in \eqref{directSum:restrictionToDiagonal}. Note that  $M_w$ is the set of indices such that $\lambda_{i}$ appears in the diagonal of $S_w$, and that that each $\lambda_{i}$ appears in the diagonal of $S_{w}$ at most once, as we can see from the expression in \eqref{directSum:restrictionToDiagonal}. If $1^n \notin M_w$, then $S_w$ is trivially positive-semidefinite, since it is obtained by adding a positive-semidefinite diagonal matrix $Y(w)$ to a positive-semidefinite matrix $Q_{0,1/\sqrt{2},\theta}^{\otimes n}(w)$. Otherwise, our appeal to Sylvester's criterion from the $0^n$ case applies again, and it is enough to prove that $\det(S_w) \geq 0$. Also, since $Q_{0,1/\sqrt{2},\theta}^{\otimes n}(w)$ has rank 1, our argument that $\det(S_{w})$ is a polynomial of minimum degree $|M_w| - 1$ applies again.

Then, using \eqref{directSum:restrictionToDiagonal}, we have that $\det(S_w)$ is given by

\begin{align}
	\epsilon^{|M_w|-1}  \(\prod_{c \in M_{w}} \dfrac{\lambda_c }{\epsilon} \) \( \epsilon - \(\dfrac{1}{2}\)^{n}  \sum_{d \in M_w}   \dfrac{g(w,d)}{\lambda_d / \epsilon}\) & 
 \end{align}

\noindent Using the recursive definition of $M_w$ in \eqref{directSum:restrictionToDiagonal}, and realizing that $1^n  \in M_w$ implies that $n_1(w)=0$, we have that 

\begin{align}
\sum_{d \in M_w}  \dfrac{g(w,d)}{\abs{\lambda_d / \epsilon}} = \( \dfrac{1}{2} \)^{n/2} (\sin(\theta) + \cos(\theta))^{n_0(w)} \(\dfrac{1}{\cos(\theta)}\)^{n_2(w)}.
\end{align}

\noindent Now, we have that

\begin{align}
 \dfrac{1}{\cos(\theta)} \leq \sin(\theta) + \cos(\theta)&  \iff   \dfrac{1}{\cos(\theta)^2}  \leq \tan(\theta) +1 \\
 & \iff   \tan(\theta)^2 \leq \tan(\theta)  \iff \theta \leq \pi/4
\end{align}

\noindent Since we are looking at the range $\theta < \theta_{n, 1/\sqrt{2}} \leq \pi/4$, and $n_0(w) + n_2(w) = n$, we have  that 

\begin{align}
(\sin(\theta) + \cos(\theta))^{n_0(w)} \(\dfrac{1}{\cos(\theta)}\)^{n_2(w)} \leq  (\sin(\theta) + \cos(\theta))^{n}.
\end{align} Therefore, since $n_2(w) \leq n$,

\begin{align}
\(\dfrac{1}{2}\)^{n}\sum_{d \in M_w} \dfrac{g(w,d)}{\lambda_d / \epsilon} & \geq   \(\dfrac{1}{2}\)^{n/2}\( 2(\cos(\theta))^n - (\cos(\theta) + \sin(\theta))^n \).
\end{align}

\noindent We see then that any $\epsilon$ that makes $\det(S_{0^n})$ non-negative will make the determinant of the other $S_w$ non-negative as well.

\subsubsection{Generalization to $\alpha \neq 1/\sqrt{2}$}

For $\alpha \neq 1/\sqrt{2}$, the changes necessary to make the proof work are limited to arithmetic adjustments. $Q_{0,\alpha,\theta}^{\otimes n}$ will now be given by

\begin{align}
&\sum_{a, c \in \{0,1\}^{n}} \ketbra{a}{c} \otimes \sum_{b, d \in \{0,1\}^{n}}  \ketbra{b}{d}  \prod_{i=0}^{n-1}   \Big( \delta_{a_i, 1-b_i}\delta_{c_i, 1- d_i}\delta_{a_i, c_i} \Big( \delta_{a_i,1}(1-\alpha^2)  +  \delta_{a_i,0}\alpha^2 \Big)   \nonumber \\
 + & \delta_{a_i,b_i}\delta_{c_i, d_i}\Big(\nonumber \delta_{a_i,1-c_i} \cdot - \alpha\sin(\theta)\sqrt{1-\alpha^2}\cos(\theta)  +    \delta_{a_i,c_i}\delta_{a_i,1}  (1-\alpha^2)  \cos(\theta)^2 \nonumber \\
  & ~~+  \delta_{a_i,c_i}\delta_{a_i,0} \alpha^2 \sin(\theta)^2   \Big) \Big). \end{align}

\noindent Note that its direct sum decomposition is not affected, since the choice of which terms of $Q_{0,\alpha,\theta}^{\otimes n}$ appear on each term does not depend on $\alpha$.

Similarly, $Y$ is given now by

\begin{align}
\sum_{a \in \{0,1\}^{n}} \lambda_{a} \ketbra{a}{a}, & \text{ where } \lambda_{a} =  \begin{cases} 
      -\epsilon (\alpha \sin(\theta))^{n-\abs{a}} \(\sqrt{1-\alpha^2}\cos(\theta)\)^{\abs{a}} & \text{for } a \neq 1^n  \\
      \epsilon \(\sqrt{1-\alpha^2}\)^{n} \cos(\theta)^n & \text{for } a=1^n  
   \end{cases} \nonumber \\
& \text{and }  \epsilon =2\(\sqrt{1-\alpha^2}\cos(\theta)\)^n - (\sqrt{1-\alpha^2}\cos(\theta) + \alpha\sin(\theta))^n.
\end{align}

As for the feasibility of $Y$, we have then that $\det(S_w)$  is given by 

\begin{align}
	\epsilon^{|M_w|-1} \( \prod_{c \in M_{w}} \dfrac{\lambda_c }{\epsilon}  \) \( \epsilon -  \sum_{d \in M_w}   \dfrac{g(w,d) \alpha^{2(n-\abs{d})} (1-\alpha^2)^{\abs{d}}}{\lambda_d / \epsilon}\) & ,
 \end{align}

\noindent again non-negative whenever

\begin{align}
\epsilon & \leq   \sum_{d \in M_w}   \dfrac{g(w,d) \alpha^{2(n-\abs{d})} (1-\alpha^2)^{\abs{d}}}{\lambda_d / \epsilon} \nonumber \\
 & =   2  \( \sqrt{1-\alpha^2} \)^{n} \cos(\theta)^{2n_0(w) - n} -  \sum_{d \in M_w}   \dfrac{g(w,d) \alpha^{2(n-\abs{d})} (1-\alpha^2)^{\abs{d}}}{\abs{\lambda_d} / \epsilon}  \label{eq:conditionOnEpsilon} .
\end{align}

Note that we have now that using the recursive definition of $M_w$ in \eqref{directSum:restrictionToDiagonal}, 

\begin{align}
 & \sum_{d \in M_w}   \dfrac{g(w,d) \alpha^{2(n-\abs{d})} (1-\alpha^2)^{\abs{d}}}{\abs{\lambda_d} / \epsilon} \nonumber \\
 = & (\alpha\sin(\theta) + \sqrt{1-\alpha^2}\cos(\theta))^{n_0(w)} \(\dfrac{ \sqrt{1-\alpha^2}}{\cos(\theta)}\)^{n_2(w)}. \nonumber
\end{align}

To prove that $\eqref{eq:conditionOnEpsilon}$ holds we will need an argument slightly more involved than the corresponding one for the $\alpha=\frac{1}{\sqrt{2}}$ case. First, we consider that for $n_0(w)=n$, the right hand side of $\eqref{eq:conditionOnEpsilon}$ is equal to $\epsilon$, by definition of $\epsilon$. Then, we prove that the right hand side of  $\eqref{eq:conditionOnEpsilon}$ increases as we decrement $n_0(w)$, and increase $n_2(w)= n - n_0(w)$ in parallel. This is because the positive term in the right hand side increases with each decrease of $n_0(w)$, and it does so by a larger factor than the one by which the negative term decreases. More rigorously, consider the expression

\begin{align}
k = \dfrac{1}{\cos(\theta)^2} - \dfrac{ \sqrt{1-\alpha^2}}{\(\alpha\sin(\theta) + \sqrt{1-\alpha^2}\cos(\theta)\)\cos(\theta)  }  .
\end{align}

\noindent First, note that 

\begin{align}
k \geq 0 & \iff \sqrt{1-\alpha^2} \cos(\theta)^2 \leq \(\alpha\sin(\theta) + \sqrt{1-\alpha^2}\cos(\theta)\)\cos(\theta) \\
& \iff \cos(\theta) \leq \dfrac{\alpha}{\sqrt{1- \alpha^2}} \sin(\theta) + \cos(\theta) \\
& \iff 0 \leq \dfrac{\alpha}{\sqrt{1- \alpha^2}} \sin(\theta),
\end{align}

\noindent which is always true when $0 \leq \theta \leq \pi/2$, which is always the case within the trigonometric domain that we consider. Then, if we denote the right hand side of $\eqref{eq:conditionOnEpsilon}$ by $r_{n_0(w)}$, we have the recursive relation

\begin{align}
r_{n_0(w)} = r_{n_0(w) + 1} \dfrac{1}{\cos(\theta)^2} + k (\alpha\sin(\theta) + \sqrt{1-\alpha^2}\cos(\theta))^{n_0(w)}\(\dfrac{ \sqrt{1-\alpha^2}}{\cos(\theta)}\)^{n - n_0(w)} \nonumber
\end{align}

\noindent We can see indeed that this defines an increasing sequence as we decrease $n_0(w)$, since the second summand is positive, and the first summand multiplies the previous value of $r$ by an amount greater than one. We have then successfully proved that  $\eqref{eq:conditionOnEpsilon}$ holds in the $\alpha \neq \frac{1}{\sqrt{2}} $ case.

\end{proof}

%-----------------------------------------------------------------------------%
\subsection{Derivation for Theorem~7}
\label{app:proof-theorem-under-error}
%-----------------------------------------------------------------------------%

\begin{proof}

For didactic purposes, we show our derivation along the line of thought used by us when obtaining it. Therefore, we first consider simple proofs for two particular cases, and then finish with a general proof.

\subsubsection{Absence of hedging for the protocol in \cite{molina2012hedging}}

It is easy to establish that in a generalization of the example in  \cite{molina2012hedging} , the hedging behavior {\it disappears} if Bob can avoid returning an answer. This generalization considers the set of protocols where the initial quantum state shared between Alice and Bob is a pure state $\psi$ such that  $\tr_{\X}(\psi \psi^*) =  \I_{\Z}/\dim(\Z) $. It suffices to prove it for one of such states, as the other ones can be obtained from it by Bob applying a unitary. We prove it then for
\begin{align}
 \psi = \frac{1}{\sqrt{\dim(\X)}}\sum_i e_i \otimes e_i,
\end{align}

\noindent with $e_i$ being the computational basis for $\X$, and corresponding to the case $\dim(\X) = \dim(\Z)$.

The reason no hedging behavior is possible is because in this situation, it is always possible for Bob to make sure he obtains the desired outcome. To see this, notice that the operator that we apply to get $Q_a$ from $P_a$ is the identity divided by $\dim(\X)$. Similarly, $E = \I_{\X \otimes \Y}/\dim(\X)$. Therefore, $(E^+)^{1/2}Q_a(E^+)^{1/2} = P_a$. As this is a projector into a non-empty space (from the assumption that Bob has a nonzero probability of obtaining the desired outcome), the norm of this operator is $1$.

\subsubsection{Absence of hedging in the classical case}

We look now at the behavior when a game is repeated twice in parallel, and the information exchanged between Alice and Bob is classical. This is reflected in the operators $\rho$ and $P_a$ we consider in our model being diagonal matrices. As $\rho$ is a diagonal matrix, then $\Psi_{\rho}$ maps diagonal matrices to diagonal matrices, so $E$ and the $Q_a$ are diagonal too. Then, if we denote by $\Omega(E)$ the matrix that has a one in a position whenever the corresponding entry of $E$ is nonzero, and a zero otherwise, we have that
\begin{equation}
	\begin{aligned}
		\norm{\Lambda_{1,2}} = \Big \lVert \Omega(E) \otimes \Omega(E) - \left( (E^+)^{1/2} \otimes  (E^+)^{1/2} \right)  \left(Q_{1-a} \otimes Q_{1-a} \right) \left( (E^+)^{1/2} \otimes  (E^+)^{1/2} \right) 
  \Big \rVert.
	\end{aligned} \nonumber
\end{equation}

Now, whenever $\Omega(E)$ has a zero entry, $ (E^+)^{1/2} Q_{1-a} (E^+)^{1/2} $ has a zero entry as well in that position, as $Q_{1-a} \leq E$. We define now $\lambda_E(X)$ as the minimum entry of a diagonal matrix $X$, restricted to the positions where $E$ has a nonzero entry. We have then that the value of the game when Bob is trying to win one out of two parallel repetitions is given by:
\begin{equation}
	\begin{aligned}
		& ~1 - \lambda_E  \Big ( \left( Q_{1-a} \otimes Q_{1-a} \right) \left( (E^+)^{1/2} \otimes  (E^+)^{1/2} \right)    \Big(Q_{1-a} \otimes Q_{1-a} \Big)  \Big)  \\
		= & ~ 1 - \lambda_E \( (E^+)^{1/2} Q_{1-a} (E^+)^{1/2} \)^2.
	\end{aligned}
\end{equation}

Since we have that 
\begin{equation}
	\begin{aligned}
		\Omega(E) &=  (E^+)^{1/2} E (E^+)^{1/2}  \\		
		&= (E^+)^{1/2} (Q_a + Q_{1-a}) (E^+)^{1/2} \\
		&= (E^+)^{1/2} Q_{1-a} (E^+)^{1/2} + (E^+)^{1/2} Q_a (E^+)^{1/2} \\
	\end{aligned}
\end{equation}

\noindent we have then that
\begin{equation}
	\begin{aligned}
		\lambda_E \( (E^+)^{1/2} Q_{1-a} (E^+)^{1/2} \)^2 = 1 -  \norm{ (E^+)^{1/2} Q_{a}  (E^+)^{1/2}}
	\end{aligned}
\end{equation}

\noindent so

\begin{equation}
	\begin{aligned}
		 1 - \lambda_E \( (E^+)^{1/2} Q_{1-a} (E^+)^{1/2} \)^2 =  \norm{ (E^+)^{1/2} Q_{a}  (E^+)^{1/2}}. 
	\end{aligned}
\end{equation}

Therefore, there is no hedging in this case. Our argument applies similarly to the case where Bob is trying to win $k$ out of $n$ repetitions.

\subsubsection{Absence of hedging in the general case}

We begin by defining the following operators:
\begin{align}
	A = \Lambda = (E^+)^{1/2} Q_a (E^+)^{1/2},  B = (E^+)^{1/2} E (E^+)^{1/2}.
\end{align}

\noindent Note that $[Q_a, (E^+)E] = 0$, as $(E^+)E$ is equal to the identity on the support of $E$ and zero outside it, and  $Q_a \leq E$, so $E^+ E Q_a = Q_a E^+ E = Q_a$. We have then that $[A,B]=0$, so $A$ and $B$ are simultaneously diagonalizable. This means that any tensor products of $A$, $B$, and  $\I$ of the same dimension are simultaneously diagonalizable as well.

We consider first the case where $k=1$ and $n=2$, and then use a proof by induction to take care of larger $n$ and $k$. Using the operators $A$ and $B$, we can use the fact that $Q_{1-a} = E - Q_a$  to write  $\Big \lVert \Lambda_{1,2} \Big \rVert$  in terms of $A$ and $B$ as 

\begin{align}
	 \Big \lVert  A \otimes B + B \otimes A - A \otimes A \Big \rVert  \leq  \Big \lVert  A \otimes \I + \I \otimes A - A \otimes A \Big \rVert  = 2\norm{A} - \norm{A}^2,
\end{align}

\noindent where the inequality follows from the fact that $0 \leq B \leq \I$. The equality  follows from considering a basis where $A$ is diagonal, and using the fact that since $Q_a \leq E$, $0 \leq A \leq \I$, so all the eigenvalues of $A$ are at most $1$.

We have then that $ \norm{\Lambda_{1,2}} =1 - (1 - \norm{A})^2$, since the fact that Bob can just choose to play independently implies $\norm{\Lambda_{1,2}} \geq 1 - (1 - \norm{A})^2$. Therefore, we obtain that playing each game independently is an optimal behavior.

In the general case where Bob is trying to win $k$ out of $n$ games, we can again express $Q_{1-a}$ as $E - Q_a$, and thus reduce $\Lambda_{k,n}$ to a sum of tensor products of $A$ and $B$.

Consider first the case where $k=1$. Then observe that we can write 

\begin{align}
\Lambda_{1,n} = \Lambda_{1, n-1}\otimes(B-A) + B^{\otimes n-1}\otimes A  \leq  \Lambda_{1, n-1}\otimes(\I-A) + \I^{\otimes n-1}\otimes A
\end{align}

Using as basis the $n^{th}$ tensor product of a basis where $A$ is diagonal, we obtain by induction on $n$ that $\norm{\Lambda_{1,n}} = 1 - (1-\norm{A})^n$. This is because for diagonal positive semidefinite matrices $J \leq \I$ and  $K$, we have $\norm{J(\I-K) + \I \cdot K} = \norm{J}(1-\norm{K}) + \norm{K}$. 

Note as well that if  $x$ is a largest eigenvalue eigenvector of $\Lambda$, a maximum-eigenvalue eigenvector of $\Lambda_{1,n}$ is given by $x^{\otimes n}$. Using this fact, we obtain a proof for the case with $k > 1$. To do this, observe that

\begin{align}
\Lambda_{k,n}  = \Lambda_{k,n-1} \otimes(B-A) + \Lambda_{k-1,n-1} \otimes  A 
  \leq  \Lambda_{k,n-1} \otimes(\I-A) + \Lambda_{k-1,n-1} \otimes  A
\end{align}

Then, using again as basis the $n^{th}$ tensor product of a basis where $A$ is diagonal, we obtain by induction that $\norm{\Lambda_{k,n}} = 1 - \sum^{k-1}_{t = 0} {n \choose t}  \norm{A}^t (1 - \norm{A})^{n-t}$, and that for all choices of $k$ and $n$, a maximum-eigenvalue eigenvector of $\Lambda_{k,n}$ is given by $x^{\otimes n}$, for $x$ a largest eigenvector of $\Lambda$. This is because for diagonal positive semidefinite matrices $J, K, H$, where $J$ and $H$ share a largest eigenvector, and $\norm{J} \leq \norm{H}$, we have $\norm{J(\I-K) + H \cdot K} = \norm{J}(1-\norm{K}) + \norm{H}\norm{K}$.

We obtain then that in this setting, no quantum advantage can be obtained by correlating Bob's strategy between parallel repetitions.

\end{proof}

\end{document}